\documentclass[12pt,final]{article}
\usepackage{amssymb,amsmath,amsthm}
\usepackage{epsfig}
\usepackage{graphics}

\newtheorem{theorem}{Theorem}[section]
\newtheorem{lemma}[theorem]{Lemma}

\newtheorem{definition}[theorem]{Definition}

\newcommand{\SNR}{\sf SNR}
\newcommand{\SIR}{\sf SIR}
\newcommand{\SINR}{\sf SINR}

\begin{document}
\title{Spectrum Sharing between Wireless Networks}
\author{Leonard Grokop \quad David N.C. Tse\\
Department of Electrical Engineering and Computer Sciences\\
University of California\\
Berkeley, CA 94720, USA\\
\{lgrokop,dtse\}@eecs.berkeley.edu} \maketitle

\begin{abstract}
We consider the problem of two wireless networks operating on the same (presumably unlicensed) frequency band. Pairs
within a given network cooperate to schedule transmissions, but between networks there is competition for spectrum. To
make the problem tractable, we assume transmissions are scheduled according to a random access protocol where each
network chooses an access probability for its users. A game between the two networks is defined. We characterize the
Nash Equilibrium behavior of the system. Three regimes are identified; one in which both networks simultaneously
schedule all transmissions; one in which the denser network schedules all transmissions and the sparser only schedules
a fraction; and one in which both networks schedule only a fraction of their transmissions. The regime of operation
depends on the pathloss exponent $\alpha$, the latter regime being desirable, but attainable only for $\alpha>4$. This
suggests that in certain environments, rival wireless networks may end up naturally cooperating. To substantiate our
analytical results, we simulate a system where networks iteratively optimize their access probabilities in a greedy
manner. We also discuss a distributed scheduling protocol that employs carrier sensing, and demonstrate via
simulations, that again a near cooperative equilibrium exists for sufficiently large $\alpha$.
\end{abstract}

\section{Introduction}
The recent proliferation of networks operating on unlicensed bands, most notably 802.11 and Bluetooth, has stimulated
research into the study of how different systems competing for the same spectrum interact. Communication on unlicensed
spectrum is desirable essentially because it is free, but users are subject to random interference generated by the
transmissions of other users. Most research to date has assumed devices have no natural incentive to cooperate with one
another. For instance, a wireless router in one apartment is not concerned about the interference it generates in a
neighboring apartment. Following from this assumption, various game-theoretic formulations have been used to model the
interplay between neighboring systems \cite{spectrum1}, \cite{spectrum2}, \cite{spectrum3}, \cite{spectrum4},
\cite{spectrum5}. An important conclusion stemming from this body of work is that for single-stage games the Nash
Equilibria (N.E.) are typically unfavorable, resulting in inefficient allocations of resources to users. A
quintessential example is the following. Consider a system where a pair of competing links is subjected to white-noise
and all cross-gains are frequency-flat. Suppose the transmitters wish to select a one-time power allocation across
frequency subject to a constraint on the total power expended (this problem is studied in \cite{gg1}, \cite{gg2},
\cite{gg3} where it is referred to as the {\it Gaussian Interference Game}). It is straightforward to reason (via a
waterfilling argument) that the selection by both users of frequency-flat power allocations, each occupying the entire
band, constitutes a N.E.. This {\it full spread} power allocation can be extremely inefficient. Consider a symmetric
system where the cross-gains and direct-gains are equal. At high $\SNR$ each link achieves a throughput of only 1
b/s/Hz, instead of $\frac{1}{2}\log_2(1+\SNR)$ b/s/Hz, which would be obtained if the links cooperated by occupying
orthogonal halves of the spectrum. At an $\SNR$ of 30 dB, the throughput ratio between cooperative behavior and this
full spread N.E. behavior, referred to as the {\it price of anarchy}, is about 5. This example highlights an important
point in relation to single-stage games between competing wireless links: users typically have an incentive to occupy
all of the available resources.

In this work a different approach is taken. Rather than assuming total anarchy, that is, competition between {\it all}
wireless links, we instead assume competition only between wireless links belonging to {\it different networks}.
Wireless links belonging to the same network are assumed to {\it cooperate}. In short, we assume competition on the
network level, not on the link level. In a practical setting this may represent the fact that neighboring wireless
systems are produced by the same manufacturer, or are administered by the same network operator. Alternatively one may
view the competition as being between coalitions of users \cite{coal1}.

To make the problem analytically tractable but still retain its underlying mechanics, we assume each network operates
under a random-access protocol, where users from a given network access the channel independently but with the same
probability. Analysis of random access protocols provides intuition for the behavior of systems operating under more
complex protocols, as the access probability can broadly be interpreted as the average degrees of freedom each user
occupies. For the case of competition on the link level, game-theoretic research of random-access protocols such as
ALOHA have been conducted in \cite{ra1} and \cite{ra2}. In our model each network has a different density of nodes and
chooses its access probability to maximize average throughput per user. Note this access protocol is essentially
identical to one in which users select a random fraction of the spectrum on which to communicate. Thus an access
probability of one corresponds to a full spread power allocation.

We first assume all links in the system have the same transmission range and afterwards show that the results are only
trivially modified if each link is assumed to have an i.i.d. random transmission range. We characterize the N.E. of
this system for a fixed-rate model, where all users transmit at the same data rate. We show that unlike the case of
competing links, a N.E. always exists and is {\it unique}. Furthermore for a large range of typical parameter values,
the N.E. is not full spread ---nodes in at least one network occupy only a fraction of the bandwidth. We also identify
two modes, delineated by the pathloss exponent $\alpha$. For $\alpha>4$, the N.E. behavior is distinctly different than
for $\alpha<4$ and possesses pseudo-cooperative properties. Following this we show that the picture for the
variable-rate model, in which users individually tailor their transmission rates to match the instantaneous channel
capacity, remains unchanged. Before concluding we present simulation results for the behavior of the system when the
networks employ a greedy algorithm to optimize their throughput, operating under both a random access protocol, and a
carrier sensing protocol.

In section II we formulate the system model explicitly. In section III.A we introduce the random access protocol and
analyze its N.E. behavior in the fixed-rate model. In section III.B we analyze the variable-rate model. In section IV
we extend our results to cover the case of variable transmission ranges. Section V presents simulation results and the
carrier sensing protocol. Section VI summarizes and suggests extensions. Section VII contains proofs of the main
theorems presented.

\section{Problem Setup}
Consider two wireless networks consisting of $n\lambda_1$ and $n\lambda_2$ tx-rx pairs, respectively. Without loss of
generality we will assume $\lambda_1 \le \lambda_2$. The transmitting nodes are uniformly distributed at random in an
area of size $n$. To avoid boundary effects suppose this area is the surface of a sphere. Thus $\lambda_i$ is the
density of transmitters (or receivers) in network $i$. For each transmitter, the corresponding receiver is initially
assumed to be located at a fixed range of $d$ meters with uniform random bearing. Time is slotted and all users are
assumed to be time synchronized.

Both networks operate on the same band of (presumably unlicensed) spectrum and at each time slot a subset of tx-rx
pairs are simultaneously scheduled in each network. When scheduled a tx-rx pair uses all of the spectrum. It is
generally desirable to schedule neighboring tx-rx pairs in different time slots. This scheduling model is a form of
TDM, but is more or less analogous to an FDM model where each tx-rx pair is allocated a subset of the spectrum
(typically overlapping in some way with other tx-rx pairs in the network).

Transmitting nodes are full buffer in that they always have data to send. Transmissions are assumed to use Gaussian
codebooks and interference from other nodes is treated as noise. Initially we analyse the model where all transmissions
in network $i$ occur at a common rate of $\log(1+\beta_i)$. We refer to $\beta$ as the {\it target} $\SINR$. Thus a
transmission in network $i$ is successful iff ${\SINR} > \beta_i$. Later we explore the model where transmission rates
are individually tailored to match the instantaneous capacities of the channels. The signal power attenuates according
to a power law with {\it pathloss exponent} $\alpha > 2$. We assume a high-$\SNR$ or {\it interference limited}
scenario where the thermal noise is insignificant relative to the received power of interfering nodes and thus refer to
the $\SIR$ as the $\SIR$. For a given realization of the node locations the time-averaged throughput achieved by the
$j$th tx-rx pair in network $i$ is then
\begin{equation*}
{\overline R}_j = f_j\mathbb P({\SIR_j(t)}>\beta_i)\log(1+\beta_i)
\end{equation*}

\noindent per complex d.o.f., where $f_j$ is the fraction of time the $j$th tx-rx pair is scheduled. The average
(represented by the bar above the $R$) is essentially taken over the distribution of the interference as at different
times different subsets of transmitters are scheduled.

As for $\alpha > 2$ the bulk of the interference is generated by the strongest interferer, to make the problem
tractable, we compute the $\SIR$ as the receive power of the desired signal divided by the receive power of the {\it
nearest} interferers signal. We refer to this as the {\it Dominant Interferer} assumption. Denote the range of the
nearest interferer to the $j$th receiver at time $t$ by $r_j(t)$. Then
\begin{equation*}
{\SIR}_j(t) = \frac{d^{-\alpha}}{r_j^{-\alpha}(t)}
\end{equation*}

\noindent The metric of interest to each network is its expected time-averaged rate per user,
\begin{equation*}
{\mathbb E}{\overline R} = {\mathbb E}_g \left[ f_j\mathbb P({\SIR_j(t)}>\beta_i)\log(1+\beta_i) \right]
\end{equation*}

\noindent The subscript $g$ indicates this expectation is taken over the geographic distribution of the nodes. As the
setup is statistically symmetric, this metric is equivalent to the expected {\it sum rate} of the system, divided by
$n\lambda_i$, in the limit $n\rightarrow \infty$.

\section{Random Access protocol}
\subsection{Fixed-Rate model}
Suppose each network uses the following random access protocol. At each time slot each link is scheduled i.i.d. with
probability $p_i$. The packet size is $\log(1+\beta_i)$ for all communications in network $i$. The variables $\beta_i$
are optimized over.

Let us first compute the optimal access probability for the case of a single network operating in isolation on a
licensed band, as a function of the node density and the transmission range. This problem has recently been studied
independently in \cite{jindal1}-\cite{jindal4} with equivalent results derived. In \cite{baccelli} similar results are
derived for the case where the $\SIR$ is computed based on all interferers, not just the nearest.

Let the r.v. $N_j(x)$ denote the number of interfering transmitters within range $x$ of the $j$th receiver.
\begin{align*}
{\mathbb E}{\overline R} &= {\mathbb E}_g \left[ f_j\mathbb P(r_j(t)>\beta_i^{1/\alpha}d)\log(1+\beta_I) \right] \\
&= p_i\sum_{k=1}^{n\lambda_i-1} {\mathbb P}(N_j(\beta^{1/\alpha}d) = k)(1-p_i)^k\log(1+\beta_i) \\
&= p_i\sum_{k=1}^{n\lambda_i-1} \left(\frac{\pi\beta_i^{2/\alpha}d^2}{n}\right)^k
\left(1-\frac{\pi\beta_i^{2/\alpha}d^2}{n}\right)^{n\lambda_i-k-1} \\
&\quad\quad\quad\quad\quad \times \binom{n\lambda_i-1}{k}(1-p_i)^k\log(1+\beta_i)\\
&= p_i\left(1-p_i\frac{\pi\beta_i^{2/\alpha}d^2}{n}\right)^{n\lambda_i-1}\log(1+\beta_i) \\
&\rightarrow p_i \log(1+\beta_i) e^{-\pi\lambda_id^2p_i\beta_i^{2/\alpha}}
\end{align*}

\noindent in the limit $n\rightarrow \infty$.

In order to obtain better insight into the problem at hand, a change of variables is required. We refer to the set of
all points within the transmission range as the {\it transmission disc}. The quantity $\pi\lambda_i d^2$ is the {\it
average number of nodes (tx or rx) per transmission disc}. We often refer to it simply as the {\it number of nodes per
disc} and represent it by the symbol
\begin{equation*}
N_i \triangleq \pi\lambda_i d^2.
\end{equation*}

Assume $N_i$ is larger than a certain threshold (we make this precise later). Maximizing over the access probability
yields
\begin{equation}\label{eqn:one_network_thpt}
{\mathbb E}{\overline R} \rightarrow \frac{\log(1+\beta_i)}{N_i \beta_i^{2/\alpha}} e^{-1}
\end{equation}

\noindent with the optimal access probability being
\begin{equation}\label{eqn:p_star} p_i^* =
\frac{1}{N_i\beta_i^{2/\alpha}}.
\end{equation}

\noindent One can further optimize over the target $\SIR$ so that $\beta_i$ is replaced by $\beta_i^*$ in the above two
equations. Inspection of equation (\ref{eqn:one_network_thpt}) reveals that the optimal target $\SIR$ is a function of
$\alpha$ alone. So if we define the quantity
\begin{equation*}
\Lambda_i \triangleq N_i p_i,
\end{equation*}

\noindent $\Lambda_i^*$ will be a constant, independent of $N_i$. The quantity $\Lambda_i$ represents the {\it average
number of (simultaneous) transmissions per transmission disc}. We sometimes refer to it simply as the {\it transmit
density}. Whereas the domain of $p_i$ is $[0,1]$, the domain of $\Lambda_i$ is $[0,N_i]$. Thus we see that for $N_i$
sufficiently large, the access probability should be set such that an optimal number of transmissions per disc is
achieved. What is this optimal number? What is the optimal target $\SIR$?

For the purposes of optimizing equation (\ref{eqn:one_network_thpt}), define the function $\Lambda^*(\alpha)$ as the
unique solution of the following equation
\begin{equation}\label{eqn:Lambda_star}
\frac{\alpha}{2} = \left( 1+ {\Lambda^*}^{\alpha/2} \right) \log\left(1+\frac{1}{{\Lambda^*}^{\alpha/2}}\right).
\end{equation}

\begin{figure}
\centering
\includegraphics[width=420pt]{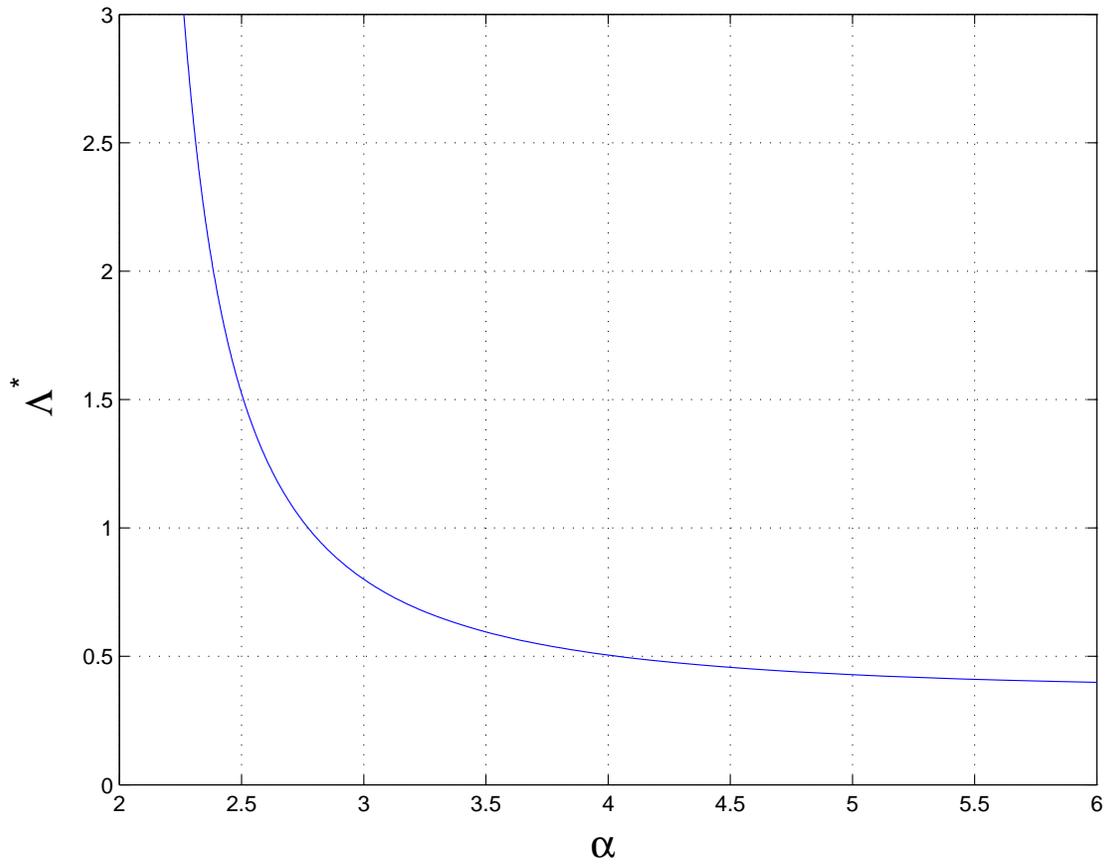}
\caption{Optimal average number of transmissions per transmission disc as a function of the pathloss
exponent.}\label{fig:Lambda_star}
\end{figure}

\noindent A plot of $\Lambda^*(\alpha)$ is given in figure \ref{fig:Lambda_star}. So as to avoid confusion, note that
the symbol $\Lambda^*$ represents a pre-defined function, not necessarily the same as the symbol $\Lambda_i^*$, which
is a variable. As equation (\ref{eqn:one_network_thpt}) is smooth and continuous with a unique maxima, by setting it's
derivative to zero we find that the optimal target $\SIR$ is $\beta^* = {\Lambda^*}^{-\alpha/2}$ and the optimal number
of transmissions per disc is $\Lambda_i^*=\Lambda^*$, when $N_i$ is larger than a certain threshold.

When $N_i$ is smaller than this threshold, there aren't enough tx-rx pairs to reach the optimal number of transmissions
per disc, even when all of them are simultaneously scheduled. In this case the solution lies on the boundary with
$p_i^*=1$. This corresponds to the scenario where the transmission range is short relative to the node density such
that tx-rx pairs function as if in isolation. It is intuitive that in this case all transmissions in the network will
be simultaneously scheduled. Our discussion is summarized in the following theorem.

\begin{theorem}\label{thm:op_access_prob} (Optimal Access Probability)
For a single network operating in isolation under the random access protocol, when $N_i > \Lambda^*$ the optimal access
probability is $p_i^* = \frac{\Lambda^*}{N_i}$, where $\Lambda^*$ is given by the unique solution of equation
(\ref{eqn:Lambda_star}). The optimal target $\SIR$ is $\beta_i^* = {\Lambda^*}^{-\alpha/2}$.

When $N_i \le \Lambda^*$, $p_i^*=1$ and $\beta_i^*$ is given by the unique solution to
\begin{equation*}
\frac{\alpha}{2N_i {\beta_i^*}^{2/\alpha}} = \left( 1+ \frac{1}{\beta_i^*} \right) \log(1+\beta_i^*).
\end{equation*}
\end{theorem}

\noindent The region satisfying $N_i > \Lambda^*$ is referred to as the {\it partial} reuse regime. The complement
region is referred to as the {\it full reuse} regime. Note the optimal access probability of the above theorem is
equivalent to the results of section IV.B in \cite{jindal1}, and those discussed under the title {\it ``Maximum
Achievable Spatial Throughput and TC''} on page 4137 of \cite{jindal3}.

Now we perform the same computation for the case where both networks operate on the same unlicensed band. In this case
there is both {\it intra-network} and {\it inter-network} interference. It is straightforward to extend the above
analysis to show that for network $i$
\begin{equation*}
{\mathbb E}{\overline R}_i \rightarrow \frac{\Lambda_i}{N_i} \log(1+\beta_i)
e^{-(\Lambda_1+\Lambda_2)\beta_i^{2/\alpha}}
\end{equation*}

\noindent in the limit $n \rightarrow \infty$. For a given $\Lambda_2$ the first network can optimize $\Lambda_1$ and
$\beta_1$, and vice-versa. That is each network can iteratively adjust its access probability and target $\SIR$ in
response to the other networks. In this sense a game can be defined between the two networks. A strategy for network
$i$ is a choice of $\Lambda_i\in [0,N_i]$ and $\beta_i>0$. Its payoff function (also referred to as {\it utility
function}) is the limiting form of ${\mathbb E}{\overline R}_i$ times $N_i$,
\begin{align}
\label{eqn:utility_fns1}
U_1\left((\Lambda_1,\beta_1),(\Lambda_2,\beta_2)\right) &= \Lambda_1 \log(1+\beta_1) e^{-(\Lambda_1+\Lambda_2)\beta_1^{2/\alpha}} \\
\label{eqn:utility_fns2} U_2\left((\Lambda_1,\beta_1),(\Lambda_2,\beta_2)\right) &= \Lambda_2 \log(1+\beta_2)
e^{-(\Lambda_1+\Lambda_2)\beta_2^{2/\alpha}}.
\end{align}

\noindent  Here we have scaled the throughput by $N_i$ to emphasize the simple form of the payoff functions. At first
glance this setup seems desirable but there is a redundancy in the way the strategy space has been defined. The problem
is that the variable $\beta_i$ only appears in $U_i$ and thus should be optimized over separately rather than being
included as part of the strategy. This leads to the following game setup.

\begin{definition}\label{def:NE} (Random Access Game)
A strategy for network $i$ in the Random Access Game is a choice of $\Lambda_i \in [0,N_i]$. The payoff functions are
\begin{align*}
U_1(\Lambda_1,\Lambda_2) &= \max_{\beta_1 > 0} \Lambda_1 \log(1+\beta_1) e^{-(\Lambda_1+\Lambda_2)\beta_1^{2/\alpha}} \\
U_2(\Lambda_1,\Lambda_2) &= \max_{\beta_2 > 0} \Lambda_2 \log(1+\beta_2) e^{-(\Lambda_1+\Lambda_2)\beta_2^{2/\alpha}}.
\end{align*}
\end{definition}

The above formulation is intuitively appealing as a networks choice of access probability constitutes its entire
strategy. If we could explicitly solve the maximization problems, the variables $\beta_i$ would be removed altogether.
When $\Lambda_1$ and $\Lambda_2$ are large this can be done and
\begin{align}
\label{eqn:appr_uf1}
U_1(\Lambda_1,\Lambda_2) &\approx \frac{\Lambda_1}{(\Lambda_1+\Lambda_2)^{\alpha/2}} \cdot (\alpha/2)^{\alpha/2}e^{-\alpha/2} \\
\label{eqn:appr_uf2} U_2(\Lambda_1,\Lambda_2) &\approx \frac{\Lambda_2}{(\Lambda_1+\Lambda_2)^{\alpha/2}} \cdot
(\alpha/2)^{\alpha/2}e^{-\alpha/2},
\end{align}

\noindent but in general it is not possible. Instead, since we are only interested in analyzing the Nash equilibrium
(N.E.) or equilibria of this game, we do the following.

Observe that the objective function within the maximization is smooth and continuous. This enables the order of
maximization to be swapped. That is, for a given $\Lambda_2$, we first maximize over $\Lambda_1$ in equation
(\ref{eqn:utility_fns1}) and then over the $\beta_1$. Likewise for equation (\ref{eqn:utility_fns2}). The benefit of
this approach is that the maximizing $\Lambda_i$ can be explicitly expressed as a function of $\beta_i$. This was
demonstrated earlier for the single network scenario. The resulting expressions are
\begin{align*}
&U_1(\Lambda_1,\Lambda_2) \\
&= \left\{
                  \begin{array}{ll}
                    \Lambda_1\log \left(1+\Lambda_1^{-\alpha/2} \right) e^{-\Lambda_2/\Lambda_1-1}, & \hbox{$\Lambda_1 < N_1$;} \\
                    \max_{\beta_1 > 0} \log(1+\beta_1) e^{-(N_1+\Lambda_2)\beta_1^{2/\alpha}}, & \hbox{$\Lambda_1 = N_1$.}
                  \end{array}
                \right. \\ \\
&U_2(\Lambda_1,\Lambda_2) \\
&= \left\{
                  \begin{array}{ll}
                    \Lambda_2\log \left(1+\Lambda_2^{-\alpha/2} \right) e^{-\Lambda_1/\Lambda_2-1}, & \hbox{$\Lambda_2 < N_2$;} \\
                    \max_{\beta_2 > 0} \log(1+\beta_2) e^{-(N_2+\Lambda_1)\beta_2^{2/\alpha}}, & \hbox{$\Lambda_2 = N_2$.}
                  \end{array}
                \right.
\end{align*}

The set of N.E. of the above game and their corresponding values of $U_1$ and $U_2$ are identical to those of the
Random Access Game. Inspection of the above equations reveals a further simplification of the problem is at hand
---the set of N.E. of the above game are identical to the set of N.E. of the following game (though the values of $U_1$ and $U_2$ at the equilibria may be
different).
\begin{definition}\label{def:NE} (Transformed Random Access Game)
A strategy for network $i$ in the Transformed Random Access Game is a choice of $\Lambda_i \in [0,N_i]$. The payoff
functions are
\begin{align*}
U_1(\Lambda_1,\Lambda_2) &= \Lambda_1\log \left(1+\Lambda_1^{-\alpha/2} \right) e^{-\frac{\Lambda_2}{\Lambda_1}-1} \\
U_2(\Lambda_1,\Lambda_2) &= \Lambda_2\log \left(1+\Lambda_2^{-\alpha/2} \right) e^{-\frac{\Lambda_1}{\Lambda_2}-1},
\end{align*}
\end{definition}

We now analyze the N.E. of the Random Access Game by analyzing the N.E. of the Transformed Random Access Game. The
first question of interest is whether or not there exists a N.E.? It turns out a unique N.E. always exists but its
nature depends crucially on the pathloss exponent. There are two {\it modes}, $2<\alpha<4$ and $\alpha>4$. We start
with the first.

\begin{theorem}\label{thm:opt_p2} (Random Access N.E. for $2 < \alpha <4$)
For $2 < \alpha < 4$ the unique N.E. occurs at $\Lambda_1^* = N_1$, and $\Lambda_2^*$ defined by either the solution of
\begin{equation} \label{eqn:opt_p2}
N_1 = \Lambda_2\left(\frac{\alpha}{2\left(1+{\Lambda_2}^{\alpha/2}\right)\log \left(1+{\Lambda_2}^{-\alpha/2}\right)} -
1\right)
\end{equation}
\noindent or $N_1$, whichever is smaller.
\end{theorem}

\noindent The N.E. described in theorem \ref{thm:opt_p2} occurs on the boundary of the strategy space. This is because
for $2<\alpha<4$ each network tries to set its number of transmissions per disc higher than the other (see the proof of
the theorem). The equilibrium is then only attained when at least one network has maxed out and scheduled all of its
transmissions simultaneously.

The N.E. can be better understood when $N_1 \gg 1$ corresponding to the case in which transmissions span several
intermediate nodes.
\begin{theorem}\label{thm:asymp_soln} (Random Access N.E. for $2 < \alpha <4$ and $N_1 \gg 1$)
In the limit $N_1 \rightarrow \infty$ the N.E. occurs at
\begin{align*}
&(\Lambda_1^*,\Lambda_2^*) \\
&\quad\quad = \left\{
                  \begin{array}{ll}
                    (N_1,N_2), & \hbox{$N_1\le N_2\le \dfrac{2}{\alpha-2}N_1$} \\
                    \left(N_1,\dfrac{2}{\alpha-2}N_1\right), & \hbox{$\dfrac{2}{\alpha-2}N_1\le N_2$.}
                  \end{array}
                \right.
\end{align*}
\end{theorem}

\begin{figure}
\centering
\includegraphics[width=420pt]{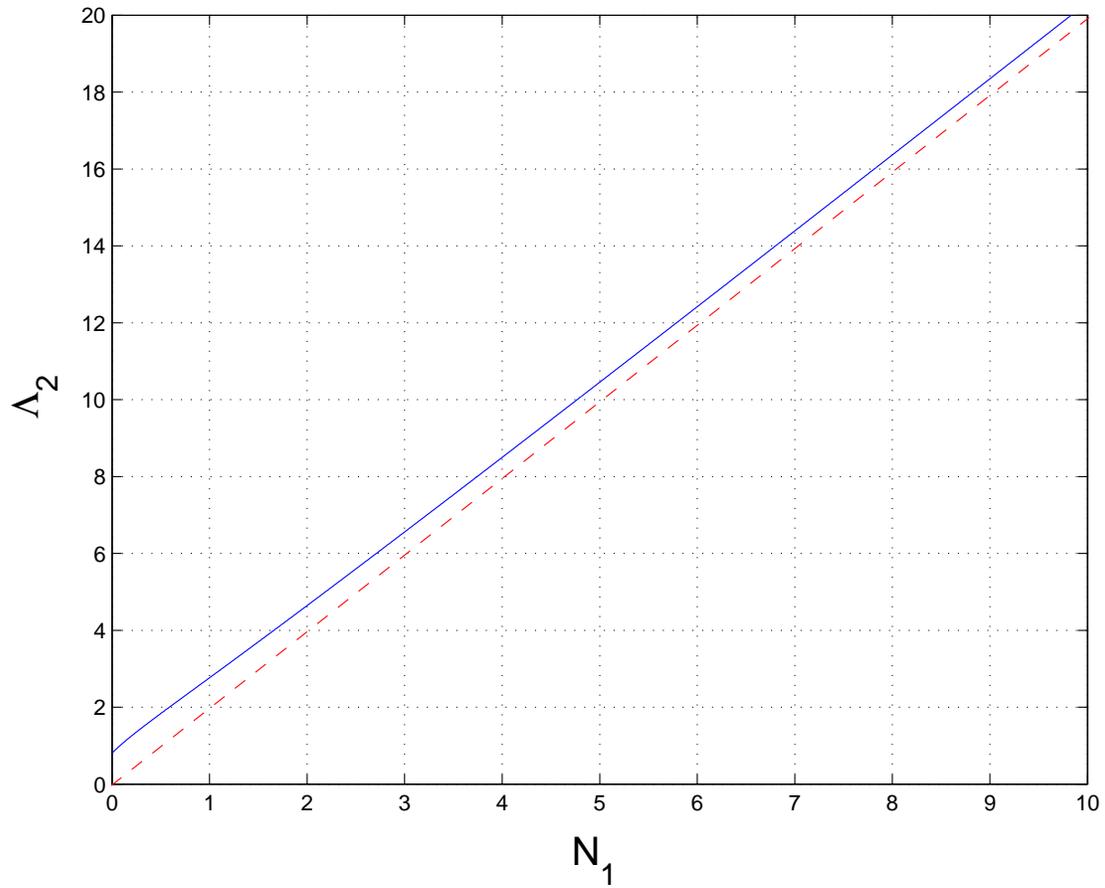}
\caption{The solid line represents the solution to equation (\ref{eqn:opt_p2}) for $\alpha = 3$. The N.E. value
$\Lambda_2^*$ is equal to the minimum of this line and $N_2$. The limiting solution used in theorem \ref{thm:apx_soln}
is plotted as a dashed line.} \label{fig:Lambda2_star}
\end{figure}

\noindent This result stems from using the limiting form $\log(1+x)\rightarrow x$ as $x\rightarrow 0$ in the utility
functions $U_1$ and $U_2$, as was done in equations (\ref{eqn:appr_uf1}) and (\ref{eqn:appr_uf2}). From it we see that
if the denser network has more than $\approx 2/(\alpha-2)$ times as many nodes as its rival, the N.E. will correspond
to partial reuse, i.e. the denser network will only occupy a fraction of the total available bandwidth. This is in
stark contrast to the case of competing {\it individual transmissions} where the N.E. typically corresponds to a full
spread, i.e. both competing links spread their power evenly across the entire bandwidth. The limit result of theorem
\ref{thm:apx_soln} is plotted in figure \ref{fig:Lambda2_star} as a dashed line.

We now investigate the average throughput at equilibrium for the mode $2<\alpha<4$. We define the metric
\begin{equation*}
U_e = U_1(\Lambda_1^*,\Lambda_2^*)+U_2(\Lambda_1^*,\Lambda_2^*)
\end{equation*}

\noindent This quantity has a natural interpretation. Recall $\mathbb E {\overline R}_i$ is the average throughput per
tx-rx pair and $N_i$ is the average number of tx-rx pairs per transmission disc in network $i$. Thus $U_i = N_i\mathbb
E {\overline R}_i$ is the {\it average throughput per transmission disc} in network $i$. This is the average number of
bits successfully received in network $i$ within an area of size $\pi d^2$ per time slot, per d.o.f.. The quantity
$U_e$ is then the average throughput per transmission disc in the system, that is, the average number of bits
successfully received {\it in both networks} within an area of size $\pi d^2$ per time slot, per d.o.f..

\begin{theorem}\label{thm:asymp_util}
In the limit $N_1\rightarrow \infty$
\begin{equation*}
U_e \rightarrow \left\{
        \begin{array}{ll}
          \dfrac{c_1(\alpha)}{N_1^{\alpha/2-1}} & \hbox{$N_1\le N_2\le \frac{2}{\alpha-2}N_1$} \\
          \dfrac{c_2(\alpha)}{(N_1+N_2)^{\alpha/2-1}}, & \hbox{$\frac{2}{\alpha-2}N_1\le N_2$.}
        \end{array}
      \right.
\end{equation*}
\noindent where $c_1(\alpha) = \left(\alpha/2-1\right)^{\alpha/2-1}(\alpha/2)e^{-\alpha/2}$ and $c_2(\alpha) =
(\alpha/2)^{\alpha/2}e^{-\alpha/2}$.
\end{theorem}

\noindent The important property of this result is that as the number of nodes per transmission disc increases, $U_e$
decreases roughly like $1/(N_1+N_2)^{\alpha/2-1}$. Let us compare this to the average throughput per transmission disc
in the cooperative case, that is when the two networks behave as if they were a single network with $N_1+N_2$ nodes per
disc. From equation (\ref{eqn:one_network_thpt}) this average throughput per disc is
\begin{equation*}
U_c = \Lambda^*\log(1+{\Lambda^*}^{-\alpha/2})
\end{equation*}

\noindent which is independent of the number of nodes per disc. Thus as the number of nodes per disc grows, so does the
price of anarchy
\begin{equation*}
\frac{U_c}{U_e} = O\left(N_1^{\alpha/2-1} \right).
\end{equation*}

\noindent For $\alpha>4$ the N.E. behavior is different. Whereas for $2<\alpha<4$ the solution always lies on the
boundary, for $\alpha>4$ it typically does not.

\begin{theorem}\label{thm:a_ge_4} (Random Access N.E. for $\alpha > 4$)
For $\alpha > 4$ the unique N.E. occurs at
\begin{equation*}
(\Lambda_1^*,\Lambda_2^*) = ( \sqrt{\Lambda^*(\alpha/2)}, \sqrt{\Lambda^*(\alpha/2)} )
\end{equation*}

\noindent if $\sqrt{\Lambda^*(\alpha/2)} < N_1$, otherwise $\Lambda_1^* = N_1$ and $\Lambda_2^*$ is defined by either
the solution of equation (\ref{eqn:opt_p2}) or $N_2$, whichever is smaller.
\end{theorem}

\noindent A plot of $\sqrt{\Lambda^*(\alpha/2)}$ versus $\alpha$ is given in figure \ref{fig:Lambda_prime}. The
condition $\sqrt{\Lambda^*(\alpha/2)} < N_1$ corresponds to network $1$ having more than $\sqrt{\Lambda^*(\alpha/2)}$
nodes per transmission disc. We refer to this as the {\it partial/partial reuse} regime.


The interpretation of theorem \ref{thm:a_ge_4} is that for $\alpha>4$ in the partial/partial reuse regime, the solution
lies in the strict interior of the strategy space. This is because on the boundary of the space network $i$ can improve
its throughput by undercutting the transmit density of network $j$, i.e. setting $\Lambda_i<\Lambda_j$. The symmetry of
the N.E. ($\Lambda_1^*=\Lambda_2^*$) then follows by observing the utility functions are symmetric and the solution is
unique.

\begin{figure}
\centering
\includegraphics[width=210pt]{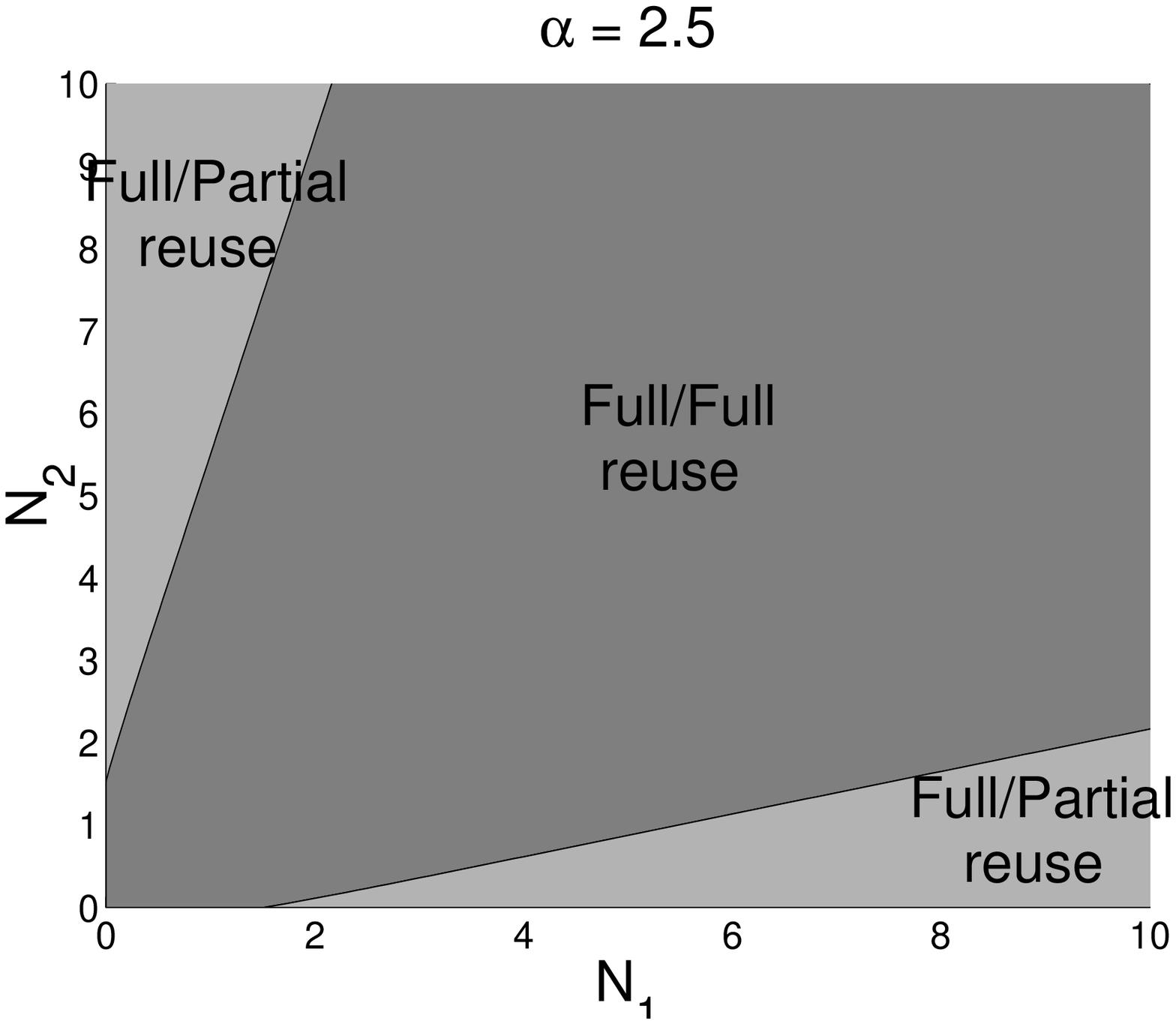}
\includegraphics[width=210pt]{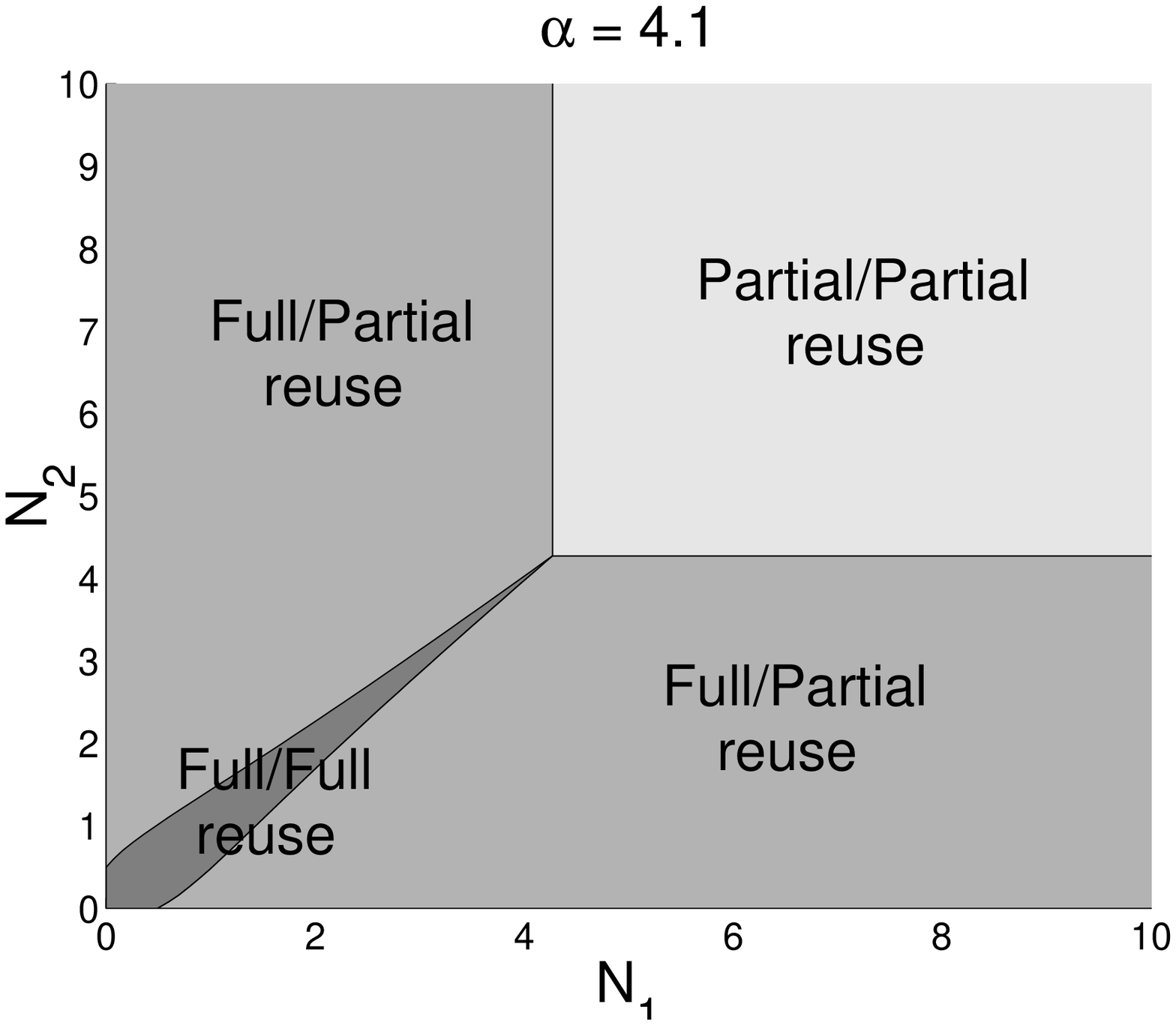}
\includegraphics[width=210pt]{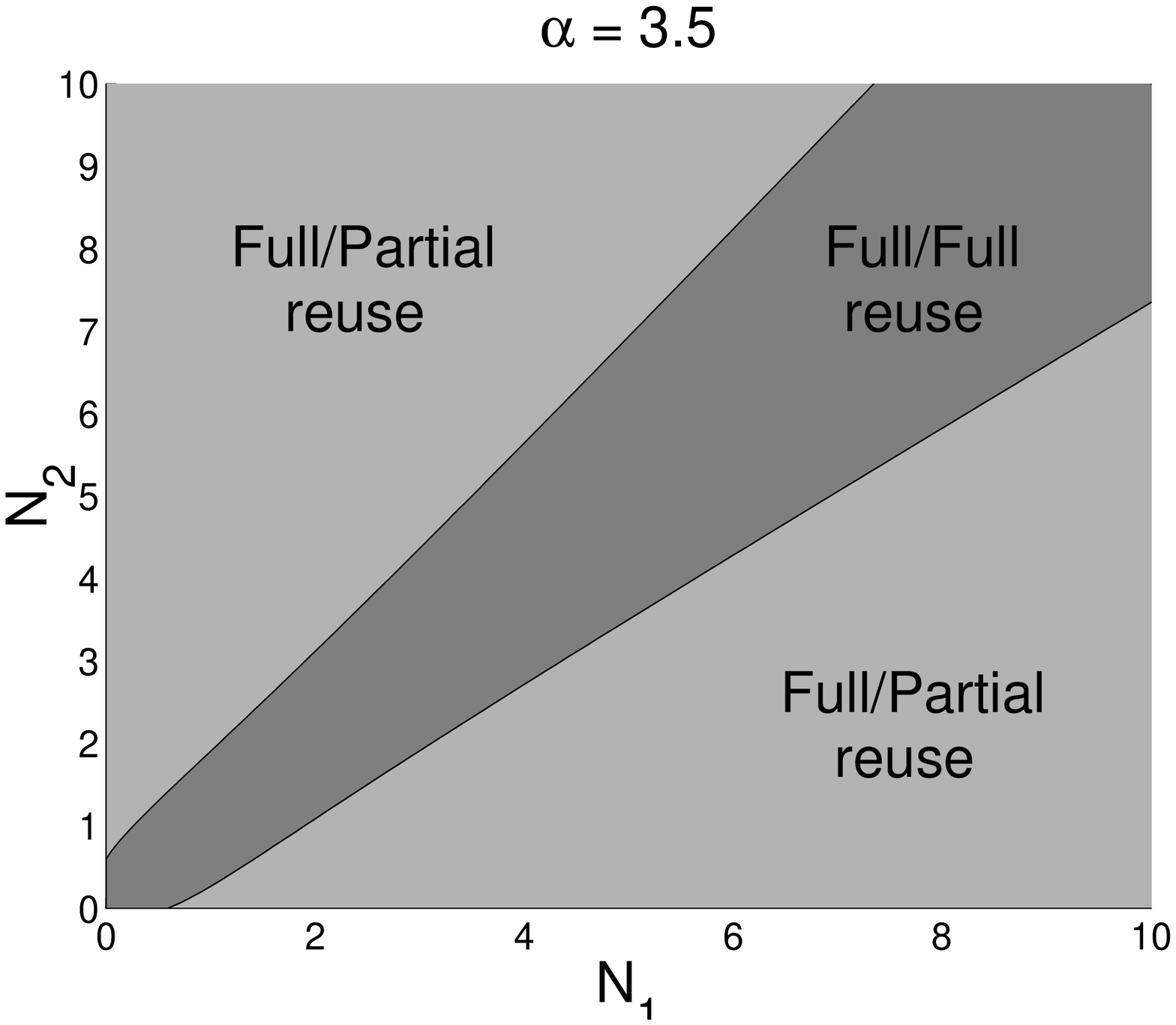}
\includegraphics[width=210pt]{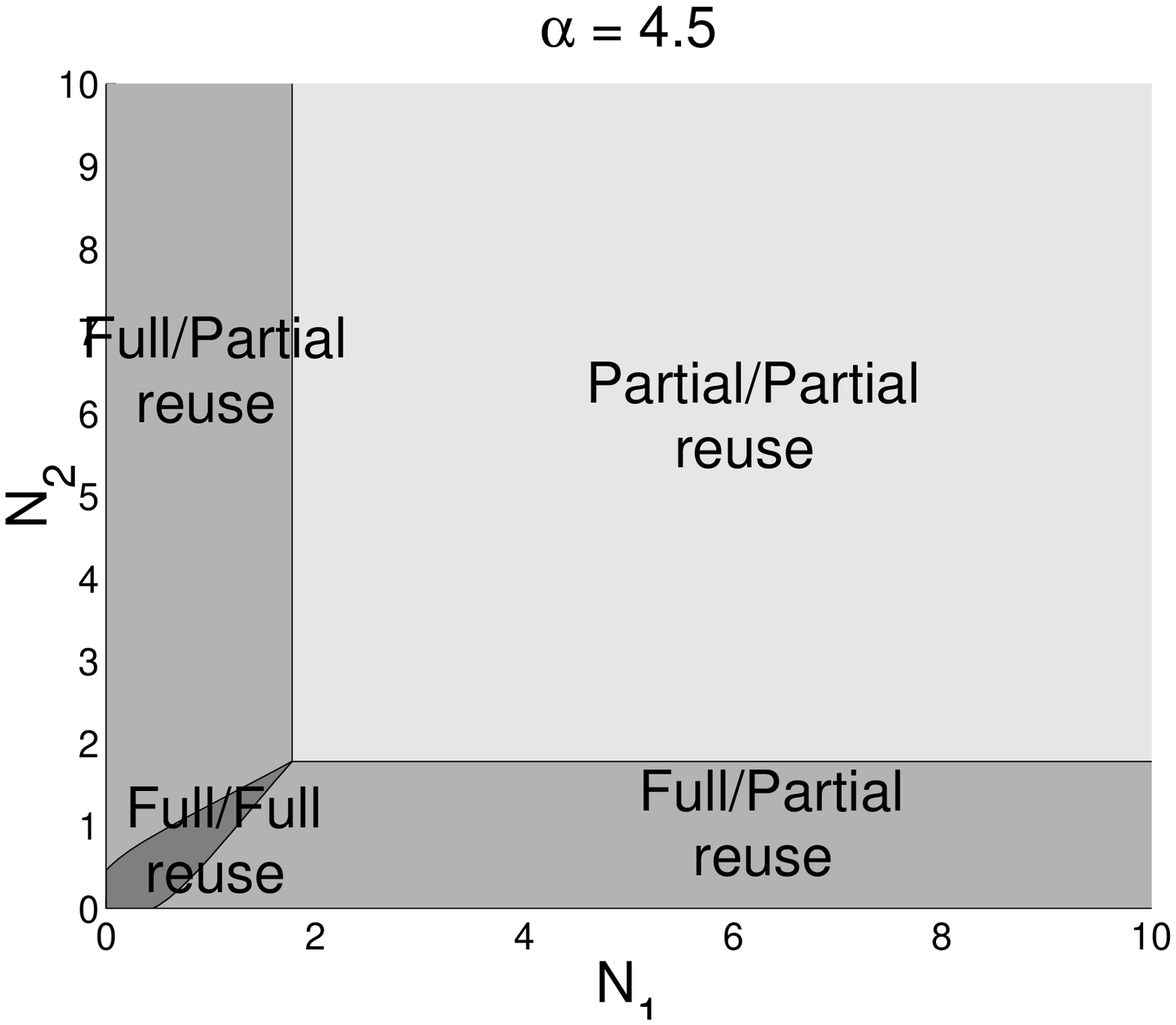}
\includegraphics[width=210pt]{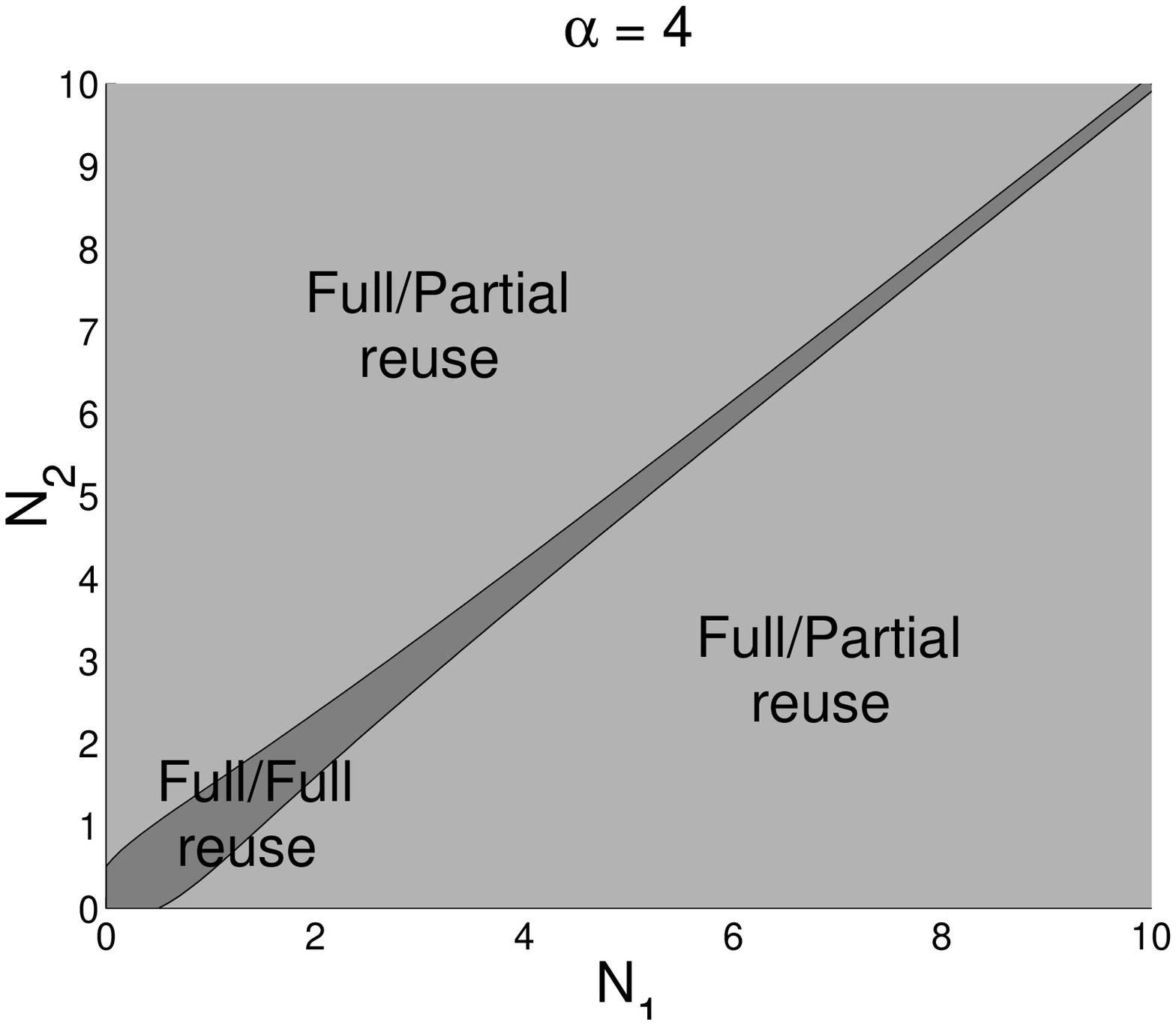}
\includegraphics[width=210pt]{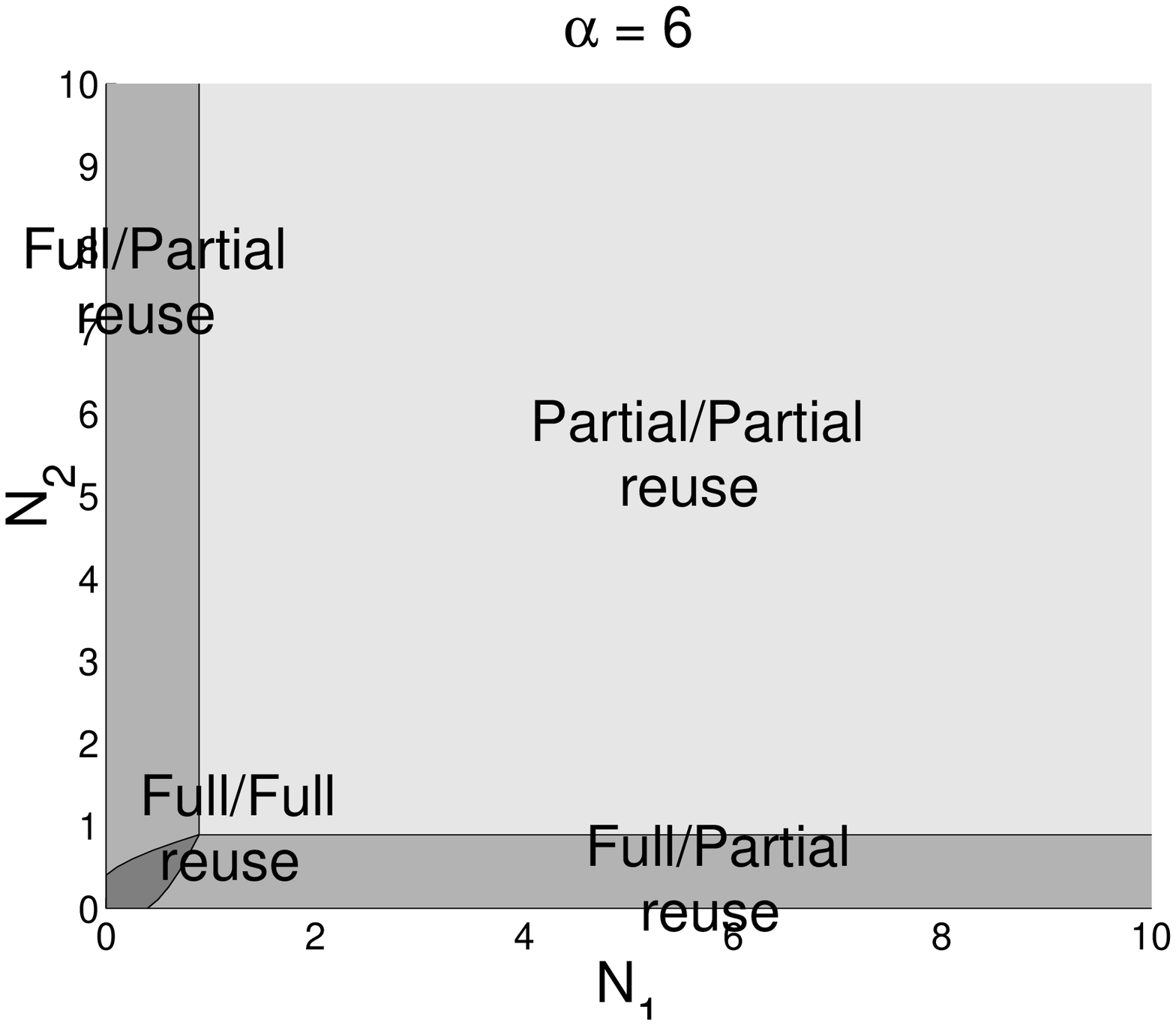}
\caption{These plots can be used to determine which regime the N.E. is in. The x-axis and y-axis corresponds to the
average number of nodes per transmission disc in network 1 and 2, respectively. Note the lower left vertex of the
partial/partial reuse regime always occurs at $(\sqrt{\Lambda^*(\alpha)},\sqrt{\Lambda^*(\alpha)})$ and the
intersection of the full/full reuse regime with the axes always occurs at $(\Lambda^*(\alpha),0)$ and
$(0,\Lambda^*(\alpha))$.} \label{fig:regimes}
\end{figure}

There is a cooperative flavor to this equilibrium in that both networks set their transmission densities to the same
level, and this level is comparable to the optimal single network density $\Lambda^*(\alpha)$. Moreover the equilibrium
level does not grow with the number of nodes per transmission disc, as it does for $2<\alpha<4$. Actual cooperation
between networks corresponds to setting the access probability based on equation (\ref{eqn:p_star}), taking into
account that the effective node density is $\lambda_1+\lambda_2$. Thus the cooperative solution is
\begin{equation*}
(\Lambda_1^*,\Lambda_2^*) = \left( \frac{\lambda_1}{\lambda_1+\lambda_2}\Lambda^*,
\frac{\lambda_2}{\lambda_1+\lambda_2}\Lambda^* \right).
\end{equation*}

\noindent Under cooperation the average throughput per transmission disc is (from equation
(\ref{eqn:one_network_thpt}))
\begin{equation*}
U_c = \frac{1}{e}\Lambda^*(\alpha)\log\left(1+\frac{1}{{\Lambda^*(\alpha)}^{\alpha/2}}\right).
\end{equation*}

\noindent Under cooperation in the partial/partial reuse regime it is
\begin{equation*}
U_e = \frac{2}{e^2}\sqrt{\Lambda^*(\alpha/2)}\log\left(1+\frac{1}{{\Lambda^*(\alpha/2)}^{\alpha/4}}\right)
\end{equation*}

\begin{figure}
\centering
\includegraphics[width=420pt]{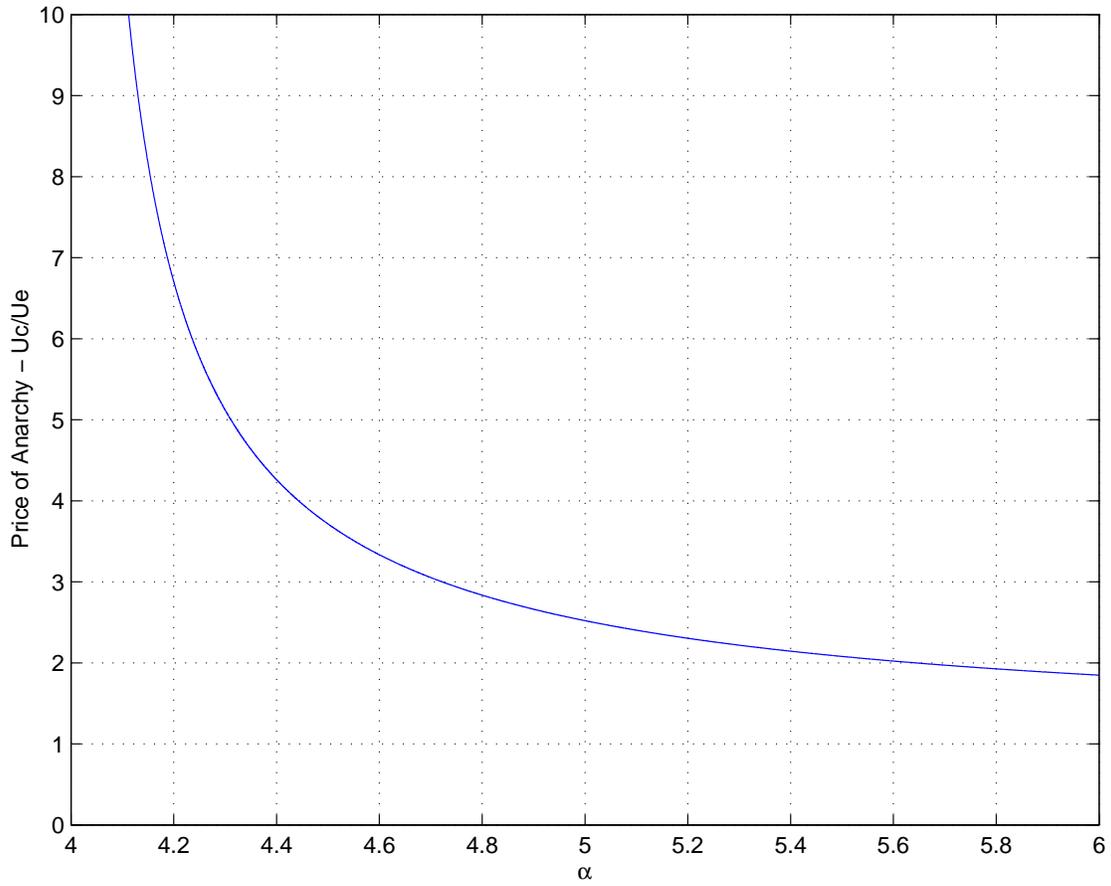}
\caption{For $\alpha>4$, the price of anarchy depends only on the pathloss exponent in the partial/partial reuse
regime. }\label{fig:poa}
\end{figure}

\noindent The price of anarchy is the ratio of these two quantities ($U_e/U_c$) and is plotted in figure \ref{fig:poa}.
Comparing the two modes we see that whereas for $2<\alpha<4$ the price of anarchy grows in an unbounded fashion with
the number of nodes per transmission disc, for $\alpha>4$ the price of anarchy in the partial/partial reuse regime is a
constant depending only on $\alpha$.

We now summarize the equilibria results. There are three regimes.
\begin{enumerate}
  \item {\it Full/Full reuse} \\
- $N_1 \le \sqrt{\Lambda^*(\alpha/2)}$ and $N_1 \le N_2\left(\alpha/2(1+{N_2}^{\alpha/2})\log
(1+{N_2}^{-\alpha/2}) - 1\right)$ \\
 - both networks schedule all transmissions
  \item {\it Full/Partial reuse} \\
- $N_1 \le \sqrt{\Lambda^*(\alpha/2)}$ and $N_1 > N_2\left(\alpha/2(1+{N_2}^{\alpha/2})\log
(1+{N_2}^{-\alpha/2}) - 1\right)$ \\
- denser network schedules all transmissions, sparser schedules only a fraction
  \item {\it Partial/Partial reuse} \\
- $N_1 > \sqrt{\Lambda^*(\alpha/2)}$ \\
- both networks schedule only a fraction of their transmissions
\end{enumerate}

\noindent In the full/full reuse regime $(\Lambda_1^*,\Lambda_2^*) = (N_1, N_2)$. In the full/partial reuse regime the
sparser network sets $\Lambda_1^* = N_1$ and the denser network sets $\Lambda_2^*$ as the solution to equation
(\ref{eqn:opt_p2}). In the partial/partial reuse regime $(\Lambda_1^*,\Lambda_2^*) = (\sqrt{\Lambda^*(\alpha/2)},
\sqrt{\Lambda^*(\alpha/2)})$.

The regimes are essentially distinguished by which boundary constraints are active. For $2 < \alpha < 4$ the
partial/partial reuse regime is not accessible. Figure \ref{fig:regimes} provides an illustrated means for determining
which regime the system is in, for a range of values of the pathloss exponent. In these plots we consider all values of
$N_1$ and $N_2$, not just those satisfying $N_1\ge N_2$. Notice that as $\alpha \rightarrow 2$ the entire region
corresponds to the full/full reuse regime, for $\alpha = 4$ almost the entire region corresponds to the full/partial
reuse regime and for $\alpha \rightarrow \infty$ the entire region corresponds to the partial/partial reuse regime.

\subsection{Variable-Rate model}
In this section we examine the case where tx-rx pairs tailor their communication rates to suit instantaneous channel
conditions, sending at rate $\log(1+{\SIR(t)})$ during the $t$th time slot. Various protocols can be used to enable
tx-rx pairs to estimate their ${\SIR}(t)$.

Consider first the single, isolated network scenario. The expected time-averaged rate per user is now
\begin{equation*}
\mathbb E{\overline R}_i = p_i\mathbb E\log(1+{\SIR}).
\end{equation*}

\noindent As before, the rate is both time-averaged over the interference and averaged over the geographic distribution
of the nodes. The $\SIR$ is the instantaneous value observed by a given rx node and is distributed according to
\begin{align*}
\mathbb P ({\SIR} > x) &= \mathbb P (r > x^{1/\alpha}d) \\
&= \left( 1 - \frac{\pi x^{2/\alpha}d^2p_i}{n} \right)^{n\lambda_i}
\end{align*}

\noindent where the variable $r$ denotes the distance to the nearest interferer. Thus
\begin{align*}
\mathbb E{\overline R}_i &= p_i\int {\mathbb P} \left( \log(1+{\SINR}) > s\right) ds \\
&= p_i\int {\mathbb P}\left( {\SINR} > x \right) \frac{dx}{1+x} \\
&= p_i\int_{0}^{\left(\frac{n}{p_i\pi d^2}\right)^{\alpha/2}} \left( 1 - \frac{\pi x^{2/\alpha}d^2p_i}{n}
\right)^{n\lambda_i} \frac{dx}{1+x} \\
&\rightarrow p_i\int_0^{\infty} e^{-\pi p_i \lambda_i d^2 x^{2/\alpha}} \frac{dx}{1+x}
\end{align*}

\noindent in the limit $n \rightarrow \infty$. Changing variables and optimizing we have
\begin{equation}\label{eqn:p_prime}
\mathbb E{\overline R}_i \rightarrow \frac{1}{N_i} \max_{0\le \Lambda_i \le 1} \Lambda_i\int_0^{\infty} e^{-\Lambda_i
x^{2/\alpha}} \frac{dx}{1+x}.
\end{equation}

\noindent Define $\Lambda'(\alpha)$ as the maximizing argument of the unconstrained version of the above optimization
problem, or more specifically as the unique solution to
\begin{equation*}
\int_0^{\infty} \frac{1 - \Lambda'x^{2/\alpha}}{1+x} e^{-\Lambda x^{2/\alpha}} dx = 0.
\end{equation*}

\begin{figure}
\centering
\includegraphics[width=420pt]{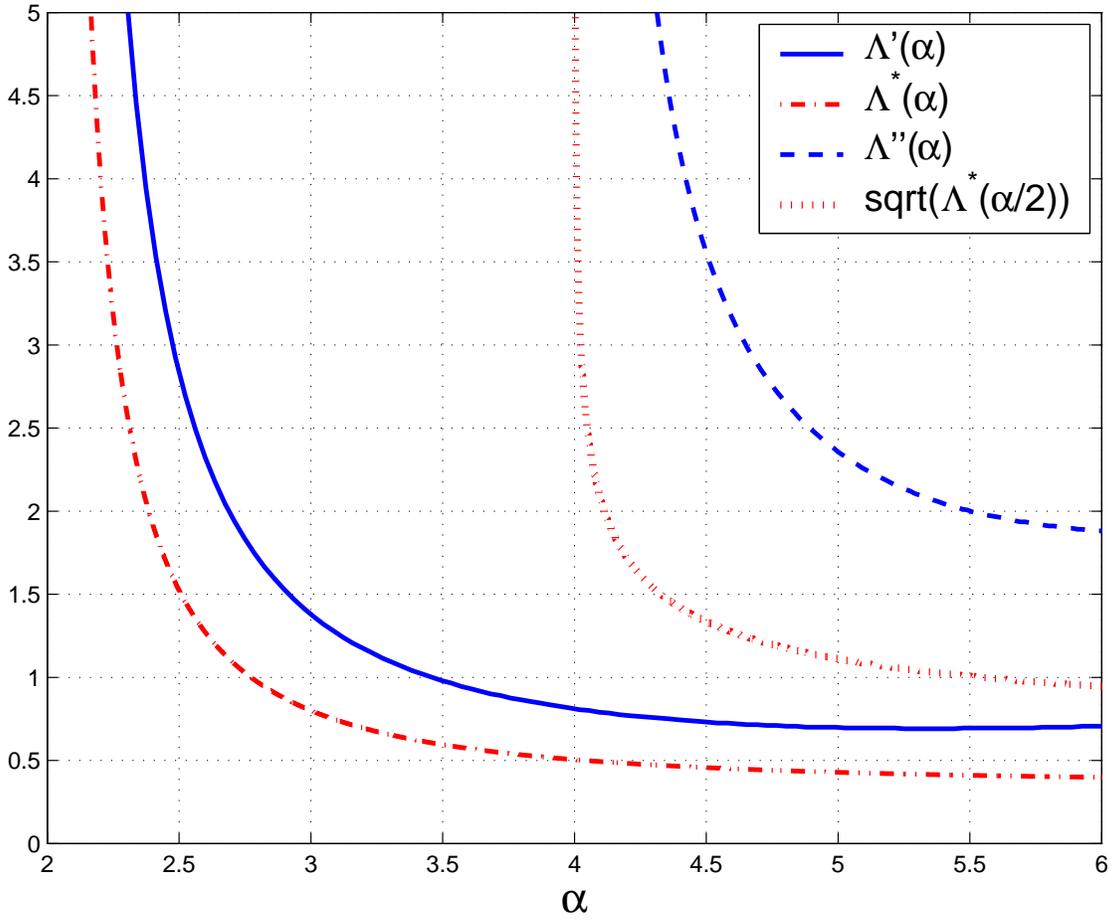}
\caption{The functions $\Lambda''(\alpha)$ and $\Lambda'(\alpha)$ for the Variable-Rate model versus the equivalent
functions $\Lambda^*(\alpha)$ and $\sqrt(\Lambda^*(\alpha/2))$ for the Fixed-Rate model.}\label{fig:Lambda_prime}
\end{figure}

\noindent The function $\Lambda'(\alpha)$ is plotted in figure \ref{fig:Lambda_prime}. Then
\begin{equation*}
p_i^* = \min(\Lambda'(\alpha)/N_i,1)
\end{equation*}

\noindent From this we see that the solution for the variable-rate case is the same as the fixed-rate solution,
differing only by substitution of the function $\Lambda'(\alpha)$ for $\Lambda^*(\alpha)$.

Now we turn to the case of two competing wireless networks. Using an approach similar to the one above it can be shown
that
\begin{equation*}
\mathbb E{\overline R}_i \rightarrow \frac{1}{N_i} \Lambda_i\int_0^{\infty} e^{-(\Lambda_1+\Lambda_2) x^{2/\alpha}}
\frac{dx}{1+x}.
\end{equation*}

\noindent In this way we can define the game between the two networks like so.
\begin{definition} (Variable Rate Random Access Game)
A strategy for network $i$ in the Variable Rate Random Access Game is a choice of $\Lambda_i\in[0,\pi\lambda_i d^2]$.
The payoff functions are
\begin{align*}
U_1(\Lambda_1,\Lambda_2) &= \Lambda_1 \int_0^{\infty} e^{-(\Lambda_1+\Lambda_2) x^{2/\alpha}} \frac{dx}{1+x} \\
U_2(\Lambda_1,\Lambda_2) &= \Lambda_2 \int_0^{\infty} e^{-(\Lambda_1+\Lambda_2) x^{2/\alpha}} \frac{dx}{1+x}.
\end{align*}
\end{definition}

From the above definition we see that the Fixed-Rate game is derived from the Variable-Rate game by merely applying a
step-function lower bound to the players utility functions, with the width of the step-function optimized, i.e.
\begin{align*}
&\int_0^{\infty} e^{-(\Lambda_1+\Lambda_2) x^{2/\alpha}} \frac{dx}{1+x} \\
&\quad\quad\quad\quad \ge \max_{\beta_i>0}
e^{-(\Lambda_1+\Lambda_2)\beta_i^{2/\alpha}}\int_{0}^{\beta_i}\frac{dx}{1+x} \\
&\quad\quad\quad\quad= \max_{\beta_i>0} \log(1+\beta_i)e^{-(\Lambda_1+\Lambda_2)\beta_i^{2/\alpha}}.
\end{align*}

\begin{figure}
\centering
\includegraphics[width=420pt]{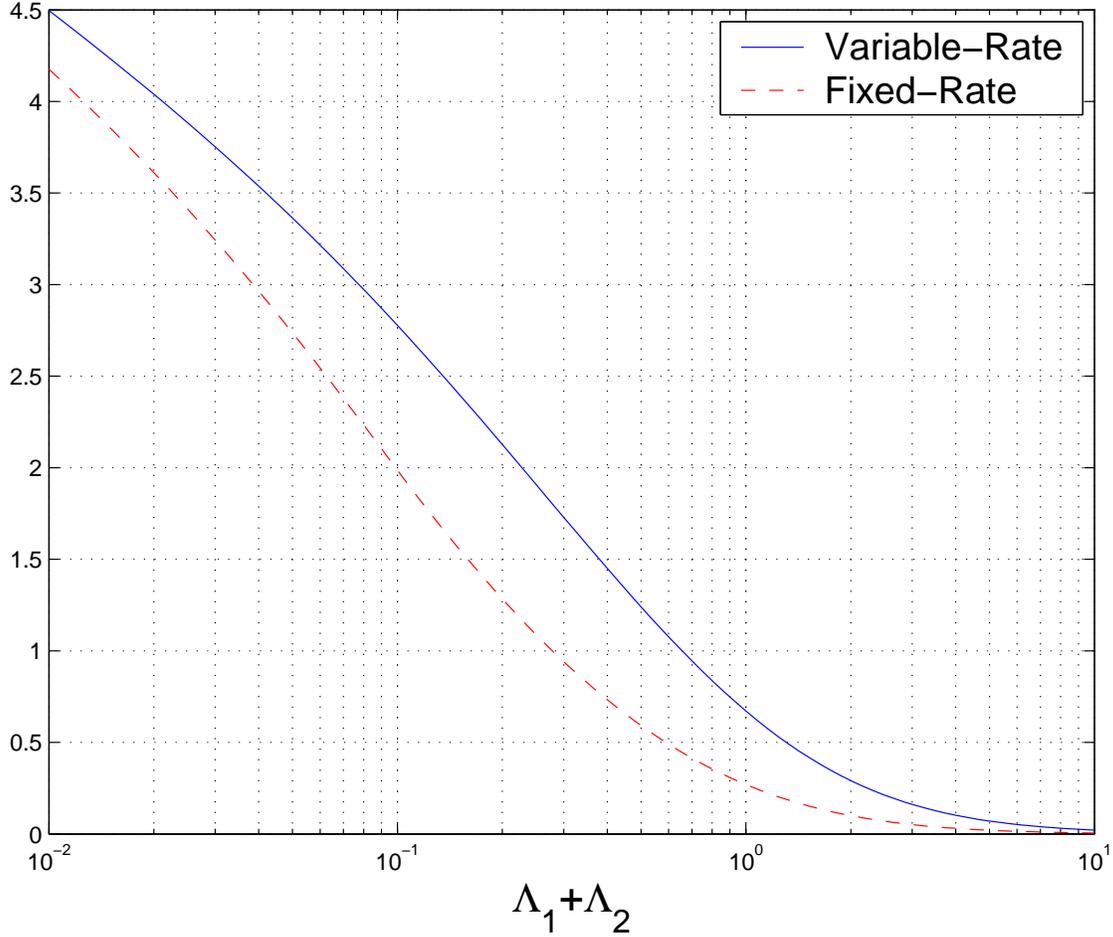}
\caption{The utility functions used in the Fixed-Rate model are lower bounds of those used in the Variable-Rate model.
In this sample plot $\alpha = 4$. The y-axis represents $U_i/\Lambda_i$. }\label{fig:vr_vs_fr}
\end{figure}

\noindent A plot comparing the expression on the left as a function of $\Lambda_1+\Lambda_2$, to the expression on the
right as a function of $\Lambda_1+\Lambda_2$, for $\alpha = 4$, is presented in figure \ref{fig:vr_vs_fr}. The figure
suggests that both expressions share a similar functional dependency on $\Lambda_1+\Lambda_2$. It is therefore natural
to wonder whether, as a consequence of this close relationship, the N.E. of the Variable-Rate game bears any
relationship to the N.E. of the Fixed-Rate game?

As in the Fixed-Rate game, the utility functions of the Variable-Rate game can be explicitly evaluated when $\Lambda_1$
and $\Lambda_2$ are large yielding
\begin{align*}
U_1(\Lambda_1,\Lambda_2) &\approx \frac{\Lambda_1}{(\Lambda_1+\Lambda_2)^{\alpha/2}}\cdot \Gamma(\alpha/2+1) \\
U_2(\Lambda_1,\Lambda_2) &\approx \frac{\Lambda_2}{(\Lambda_1+\Lambda_2)^{\alpha/2}}\cdot \Gamma(\alpha/2+1).
\end{align*}

\begin{figure}
\centering
\includegraphics[width=420pt]{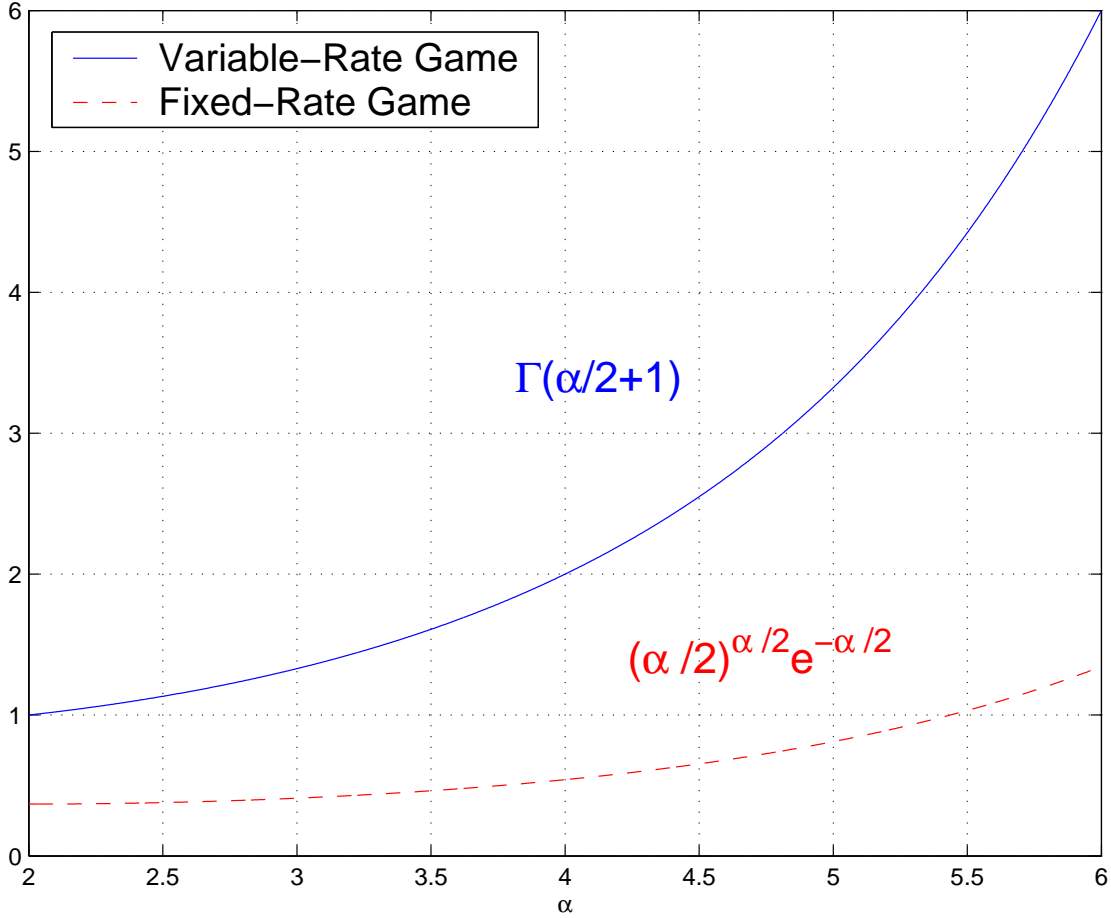}
\caption{The total system throughputs at equilibrium for the Variable-Rate and Fixed-Rate games differ only by
$\alpha$-dependent constants. These constants are plotted above. For the Variable-Rate game the constant is
$\Gamma(\alpha/2+1)$ versus $(\alpha/2)^{\alpha/2}e^{-\alpha/2}$ for the Fixed-Rate game.} \label{fig:consts}
\end{figure}

\noindent Comparing with equations (\ref{eqn:utility_fns1}) and (\ref{eqn:utility_fns2}) we see that for large
$\Lambda_1,\Lambda_2$, the utility functions of the Variable-Rate game have exactly the same functional dependency on
$\Lambda_1,\Lambda_2$ as the utility functions of the Fixed-Rate game, differing only in an $\alpha$-dependent
constant. These constants are plotted in figure \ref{fig:consts}. The plots illustrate the benefit in total system
throughput that stems from playing the Variable-rate game in place of the Fixed-Rate game.

As anticipated by the above discussion, the N.E. behavior of the Variable-Rate game parallels that of the Fixed-Rate
game. The same two modes are present, $2<\alpha<4$ and $\alpha>4$. These give rise to the same three spreading regimes,
the only difference being that the boundaries delineating them are shifted slightly. The N.E. values
$(\Lambda_1^*,\Lambda_2^*)$ in each regime take on a similar form.
\begin{theorem} \label{thm:vr_soln}
The Variable-Rate Random Access Game has a unique N.E. $(\Lambda_1^*,\Lambda_2^*)$ which lies in one of three regions.
Let $\Lambda''(\alpha)$ be the unique solution of
\begin{equation}\label{eqn:Lambda_doubleprime}
\int_0^{\infty} \frac{1 - \Lambda''x^{2/\alpha}}{1+x} e^{-2\Lambda''x^{2/\alpha}} dx = 0.
\end{equation}

\noindent for $\alpha > 4$, and equal to positive infinity for $\alpha\le 4$.
\begin{itemize}
  \item (Full/Full reuse) If $N_1 \le \Lambda''(\alpha)$ and $N_2 \le$ the unique solution over $\Lambda$ of
  \begin{equation} \label{eqn:vr_thm}
  \int_0^{\infty} \frac{1 - \Lambda x^{2/\alpha}}{1+x} e^{-(N_1+\Lambda)x^{2/\alpha}} dx = 0,
  \end{equation}
  \noindent then $(\Lambda_1^*,\Lambda_2^*) = (N_1,N_2)$.
  \item (Full/Partial reuse) If $N_1 \le \Lambda''(\alpha)$ and $N_2 >$ the unique solution of equation (\ref{eqn:vr_thm}) then
  $\Lambda_1^* = 1$ and $\Lambda_2^*$ is equal to this unique solution.
  \item (Partial/Partial reuse) If $N_1 > \Lambda''(\alpha)$ then $(\Lambda_1^*,\Lambda_2^*) = (\Lambda''(\alpha),\Lambda''(\alpha))$.
\end{itemize}
\end{theorem}

\begin{figure}
\centering
\includegraphics[width=420pt]{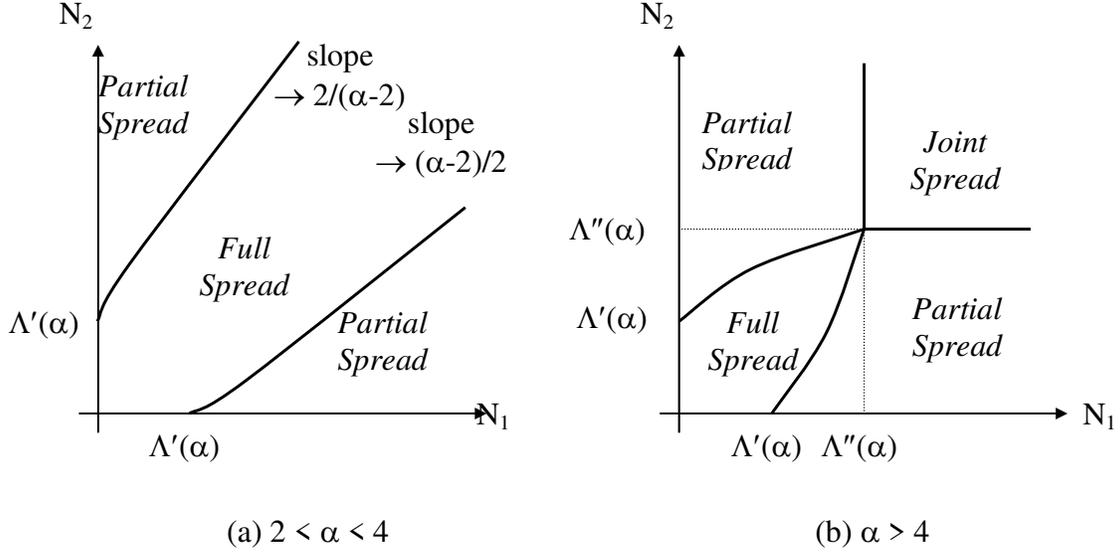}
\caption{The three regimes for the N.E. of the Variable-Rate Game.} \label{fig:vr_regimes}
\end{figure}

\noindent A regime map is provided in figure \ref{fig:vr_regimes}. As is evident from the above theorem, it is not
possible to characterize the behavior of the N.E. for the Variable-Rate game as explicitly as for the Fixed-Rate game.
This is in part due to the more complex representation of the utility functions in terms of integrals, and in part due
to the fact that the the function $\Lambda''(\alpha)$ cannot be represented in terms of the function
$\Lambda'(\alpha)$, as in the case of the Fixed-Rate game, where one function equals the square-root of the other
evaluated at $\alpha/2$.

For large $N_1$ however, we can use the approximation adopted in theorem \ref{thm:apx_soln} to explicitly characterize
the behavior of the N.E. in the full/partial reuse regime.
\begin{theorem}\label{thm:apx_soln_vr} (Variable-Rate Random Access N.E. for $2 < \alpha <4$ and $N_1 \gg 1$)
In the limit $N_1 \rightarrow \infty$ the N.E. occurs at
\begin{align*}
&(\Lambda_1^*,\Lambda_2^*) \\
&\quad\quad = \left\{
                  \begin{array}{ll}
                    (N_1,N_2), & \hbox{$N_1\le N_2\le \dfrac{2}{\alpha-2}N_1$} \\
                    \left(N_1,\dfrac{2}{\alpha-2}N_1\right), & \hbox{$\dfrac{2}{\alpha-2}N_1\le N_2$.}
                  \end{array}
                \right.
\end{align*}
\end{theorem}

\noindent Thus for $2 < \alpha < 4$ and large $N_1$, the behavior of the N.E. in the Variable-Rate game is identical to
that of the Fixed-Rate game. As discussed earlier, the values of $U_1$ and $U_2$ at equilibrium are equal to those of
the Fixed-Rate game times a constant $(\alpha/2)^{\alpha/2}e^{-\alpha/2}/\Gamma(\alpha/2+1)$.

\subsection{Explanation of Behavior}
The intuition behind our result is the following. The average throughput per link is essentially equal to the product
of the fraction of time transmissions are scheduled, and the average number of bits successfully communicated per
transmission. Adjusting the transmit density has a linear effect on the former term, but a non-linear effect on the
latter. The latter depends on the $\SIR$ and the $\SIR$ essentially depends on the pathloss exponent via
\begin{equation*}
{\SIR} \approx \left(\frac{\text{distance to transmitter}}{\text{distance to interferer}}\right)^{\alpha}
\end{equation*}

\noindent When the nearest interferer is closer than the transmitter, the ratio inside the parentheses is less than
one, and a large value of $\alpha$ substantially hurts the $\SIR$, dragging it to near zero and causing the link
capacity to drop to near zero. However, when the nearest interferer is further away than the transmitter, the ratio is
greater than one and a large value of $\alpha$ substantially improves the $\SIR$, resulting in a large link capacity.
Thus for large $\alpha$ the average number of bits successfully communicated per transmission is very sensitive to
whether or not the transmission disc is empty.

This insensitivity for sufficiently small $\alpha$ means that increasing the transmit density in network $i$ causes a
linear increase in the fraction of time transmissions are scheduled, but has little effect on the number of bits
successfully communicated per transmission, up until the point where the transmit density of network $i$ starts to
dwarf the transmit density of network $j$. Thus network $i$ will wish to increase its transmit density until it is
sufficiently larger than network $j$'s. Likewise network $j$ will wish to increase its transmit density until it is
sufficiently larger than network $i$'s. Ultimately this results in either
\begin{enumerate}
  \item a full/full reuse solution, which occurs when the sparser network max's out and winds up simultaneously scheduling
all of its transmissions, and the denser network is insufficiently dense such that its optimal transmit density based
on the sparser networks choice, results in it simultaneously scheduling all of its transmissions, or
  \item a full/partial reuse solution, which occurs when the sparser network max's out and winds up simultaneously scheduling
all of its transmissions but the denser network is sufficiently dense such that its optimal transmit density based on
the sparser networks choice, results in it simultaneously scheduling only a fraction of its transmissions.
\end{enumerate}

The opposite effect occurs for sufficiently large $\alpha$, where the average number of bits successfully communicated
per transmission depends critically on whether or not there is an interferer inside the transmission disc. In this
scenario network $i$ will set its active density to a level lower than network $j$'s, in order to capitalize on those
instances in which network $j$ happens to not schedule any transmissions nearby to one of network $i$'s receivers,
resulting in the successful communication of a large number of bits. Likewise network $j$ will set its active density
to a level lower than network $i$'s, and the system converges to the partial/partial reuse regime.

\section{Variable Transmission Range}
One of our initial assumptions was that all tx-rx pairs have the same transmission range $d$. In this section we
consider the scenario where the transmission ranges of all tx-rx pairs in the system are i.i.d. random variables $D_j$.
When the variance of $D_j$ is large, some form of power control may be required to ensure long range transmissions are
not unfairly penalized. A natural form of power control involves tx nodes setting their transmit powers such that all
transmissions are received at the same $\SNR$. This means transmit power scales proportional to $D_j^{\alpha}$. Denote
the distance from the $k$th tx node to the $j$th rx node $D_{ij}$. Then the interference power from the $k$th tx node
impinging on the $j$th rx node is proportional to $D_{kk}^{\alpha}/D_{kj}^{\alpha}$. In the fixed transmission range
scenario this quantity was proportional to $1/D_{kj}^{\alpha}$. There we assumed the bulk of the interference was
generated by the dominant interferer. Denote the scheduled set of tx nodes at time $t$ by ${\cal A}(t)$. This
assumption essentially evoked the following approximation
\begin{equation*}
\sum_{k\in\cal A(t)} 1/D_{kj}(t)^{\alpha} \approx \max_{k\in \cal A(t)} 1/D_{kj}(t)^{\alpha}.
\end{equation*}

\noindent The equivalent approximation for the variable transmission range problem is
\begin{equation*}
\sum_{k\in\cal A(t)} D_{kk}^{\alpha}/D_{kj}(t)^{\alpha} \approx \max_{k\in \cal A(t)}
D_{kk}^{\alpha}/D_{kj}(t)^{\alpha}.
\end{equation*}

\noindent Thus for variable range transmission the dominant interferer is not necessarily the nearest to the receiver.
Under this assumption the $\SIR$ at the $j$th rx node at time $t$ is then
\begin{equation*}
{\SIR}_j(t) = \frac{D_{k^*k^*}^{-\alpha}}{D_{k^*j^*}^{-\alpha}}
\end{equation*}

\noindent where $k^*$ is the index of the tx node that is closest to the $j$th receiver relative to its transmission
range.

Let us compute the throughput for the variable transmission range model under the Fixed-Rate Random Access protocol.
\begin{equation*}
{\mathbb E} {\overline R} = p_i {\mathbb P} \left( {\SIR}_j(t) > \beta_i\right)\log(1+\beta_i).
\end{equation*}

\noindent \noindent The probability the $\SIR$ is greater than the threshold
\begin{align*}
{\mathbb P} ({\SINR_j(t)> \beta_i}) &= {\mathbb P} \left( D_{kj} > \beta_i^{1/\alpha} D_{kk}, \forall k\right) \\
&= \left( \int_{0}^{\sqrt{n/\pi}/\beta_i^{1/\alpha}} {\mathbb P} \left( D_{kj} > \beta_i^{1/\alpha} x \right) {\mathbb
P}_{D_{kk}}(x)dx\right)^{n\lambda} \\
&= \left( \int_{0}^{\sqrt{n/\pi}/\beta_i^{1/\alpha}} \left(1-\frac{p_i\pi\beta_i^{2/\alpha}x^2}{n}\right) {\mathbb
P}_{D_{kk}}(x)dx\right)^{n\lambda} \\
&= \left( 1 - \frac{p_i\pi\beta_i^{2/\alpha}}{n} \int_{0}^{\sqrt{n/\pi}/\beta_i^{1/\alpha}} x^2 {\mathbb
P}_{D_{kk}}(x)dx\right)^{n\lambda} \\
&= \left( 1 - \frac{p_i\pi\beta_i^{2/\alpha}{\mathbb E}D_{kk}^2}{n}\right)^{n\lambda} \\
&\rightarrow e^{-p_i\pi \beta_i^{2/\alpha}{\mathbb E}D_{kk}^2}
\end{align*}

\noindent as $n\rightarrow \infty$. For notational simplicity let $d\equiv D_{kk}$. If we define $N_i$ as the average
number of nodes per transmission disc, where the average is taken over both the geographical distribution of the nodes
{\it and the distribution of the size of the transmission disc}, i.e.
\begin{equation*}
N_i = \pi \lambda_i {\mathbb E}_d^2
\end{equation*}

\noindent we wind up with ${\mathbb E}{\overline R} \rightarrow p_i \log(1+\beta_i) e^{-N_ip_i\beta_i^{2/\alpha}}$,
which is the same result as the fixed-transmission range model. It is straightforward to extend the analysis to the
case of two competing networks. The throughput per user in network 1 is then
\begin{equation*}
{\mathbb E}{\overline R} \rightarrow p_1 \log(1+\beta_1) e^{-(N_1p_1+N_2p_2)\beta_1^{2/\alpha}}.
\end{equation*}

\noindent Likewise for network 2. From this we see that all results for the fixed-transmission range model extend to
the variable-transmission range model by simply replacing $d^2$ by ${\mathbb E} d^2$.

\section{Simulations}
\subsection{Random Access Protocol}

In order to get a sense of the typical behavior of the players in the (Variable-Rate) Random Access game, and to
justify the validity of the Dominant Interferer assumption, we simulate the behavior of the following greedy algorithm
with the interference computed based on all transmissions in the network, not just the strongest. \\

\noindent {\bf Inputs:} $p_1(0),p_2(0),\Delta$ \\
{\bf Outputs:} ${\bf p}_i = [p_i(1),\dots,p_i(500)]$, for $i=1,2$. \\

\noindent {\bf For} $t = 1$ to $500$ \\

Form estimate $\overline{R}_1(p_1(t-1) + \Delta,p_2(t-1))$

Form estimate $\overline{R}_1(p_1(t-1) - \Delta,p_2(t-1))$ \\

{\bf If} $\overline{R}_1(p_1(t-1) + \Delta,p_2(t-1)) > \overline{R}_1(p_1(t-1) - \Delta,p_2(t-1))$

\hspace{3 mm} $p_1(t) =  \min (p_1(t-1) + \Delta,1)$

{\bf Else}

\hspace{3 mm} $p_1(t) = \max (p_1(t-1) - \Delta, 0)$

{\bf End} \\

Form estimate $\overline{R}_2(p_1(t), p_2(t-1) + \Delta)$

Form estimate $\overline{R}_2(p_1(t), p_2(t-1) - \Delta)$ \\

{\bf If} $\overline{R}_2(p_1(t), p_2(t-1) + \Delta) > \overline{R}_2(p_1(t),p_2(t-1) - \Delta)$

\hspace{3 mm} $p_2(t) =  \min( p_2(t-1) + \Delta, 1)$

{\bf Else}

\hspace{3 mm} $p_2(t) = \max(p_2(t-1) - \Delta,0)$

{\bf End} \\

\noindent {\bf End} \\

Each update time $t$, network 1 temporarily sets its access probability to $p_1(t-1)+\Delta$ and measures the resulting
throughput, averaged over 200 transmission times. This is denoted $\overline{R}_1(p_1(t-1) + \Delta,p_2(t-1))$. It then
repeats this measurement for an access probability of $p_1(t-1)-\Delta$. This is denoted $\overline{R}_1(p_1(t-1) -
\Delta,p_2(t-1))$. It then either permanently increases its access probability to $p_1(t) = p_1(t-1)+\Delta$ or
permanently decreases it to $p_1(t) = p_1(t-1) - \Delta$ depending on which option it estimates will lead to a higher
throughput. Now network 2 performs the same operation. It uses a total of 400 time slots to measure the effect of
increasing versus decresing its access probability and then either sets $p_2(t) = p_2(t-1)+\Delta$ or $p_2(t) =
p_2(t-1)-\Delta$. If $\Delta$ is small, then both networks can perform these measurement operations simultaneously
without significantly affecting the outcome.

\begin{figure}
\centering
\includegraphics[width=210pt]{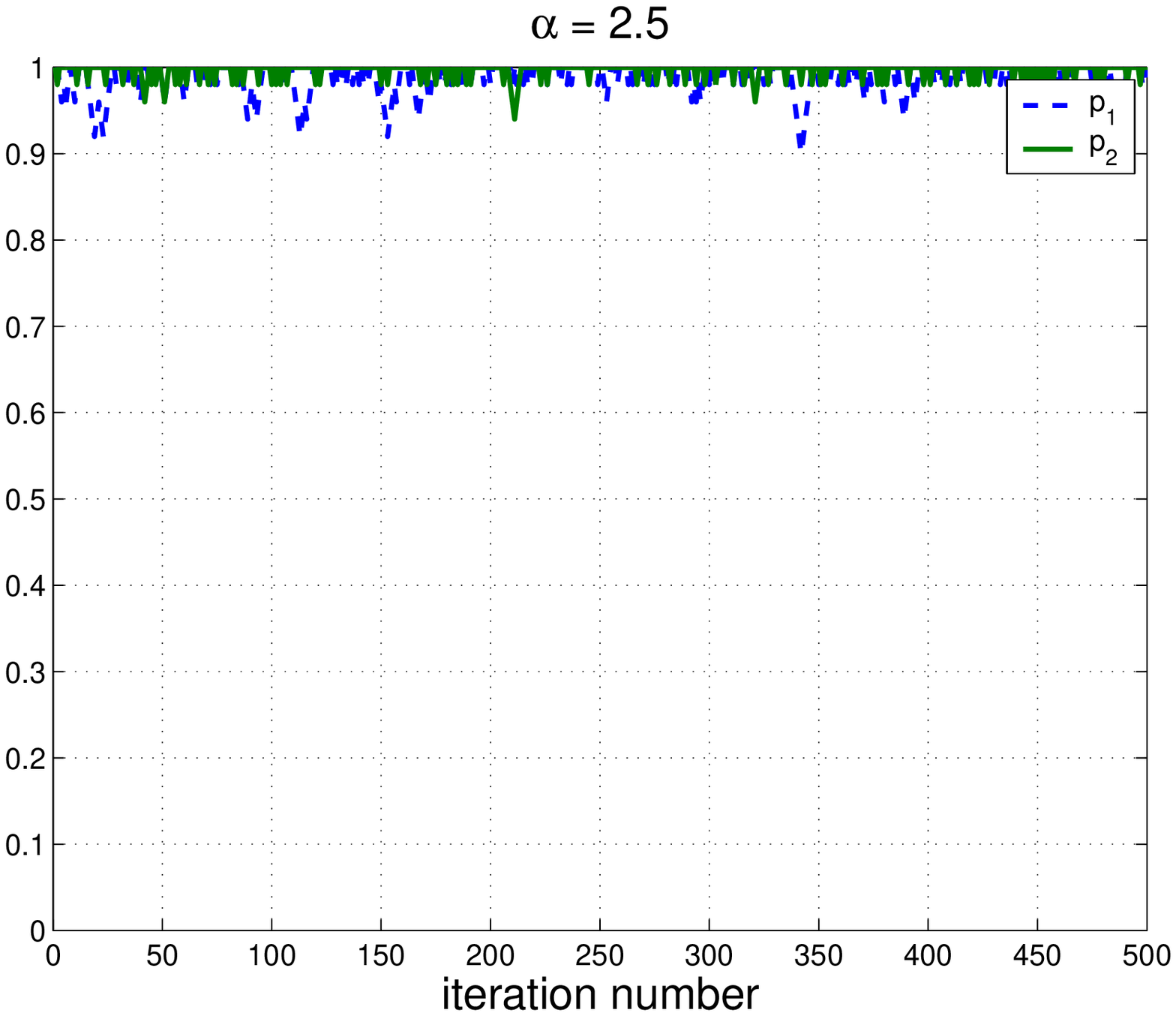}\\
\includegraphics[width=210pt]{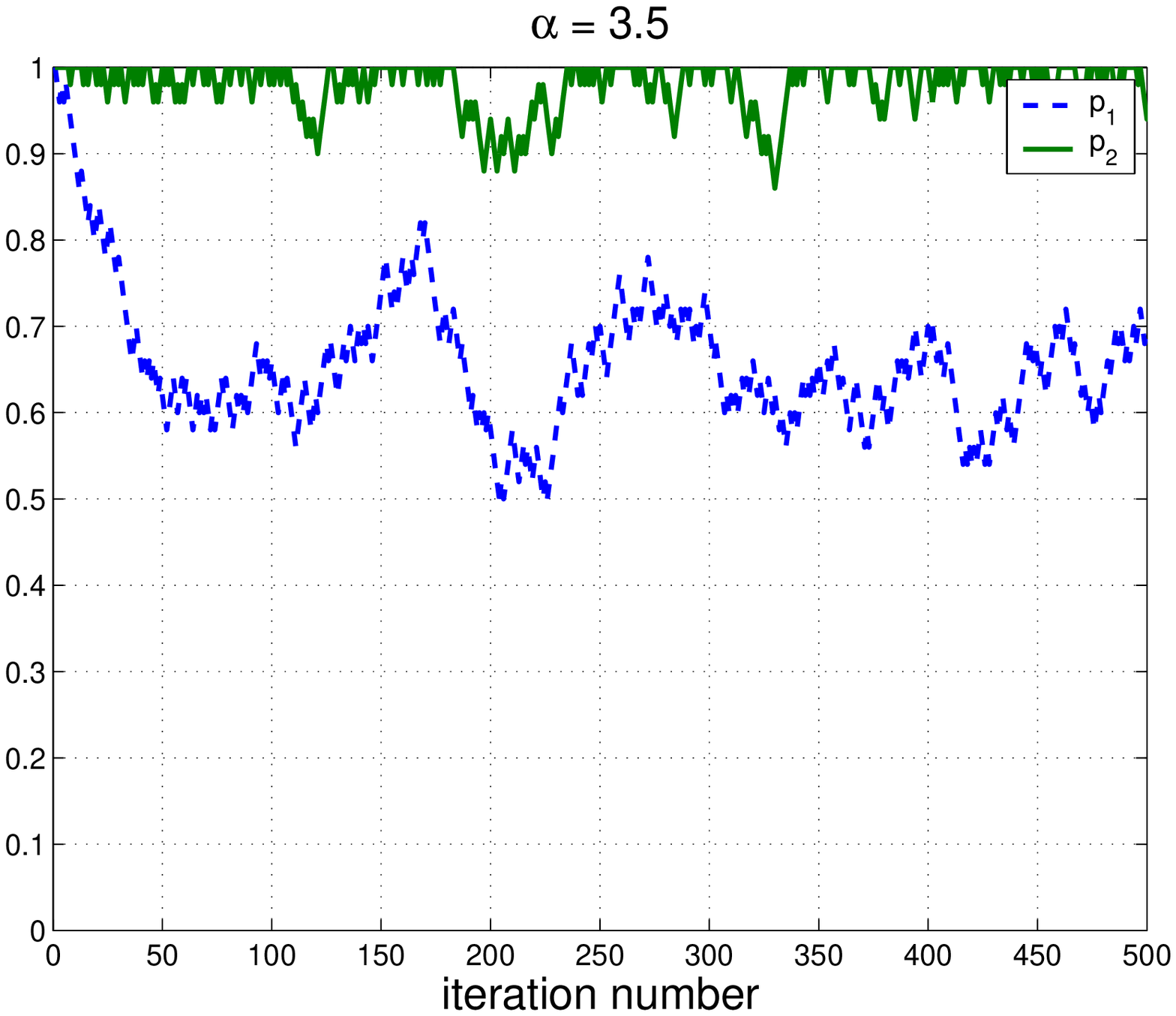}\\
\includegraphics[width=210pt]{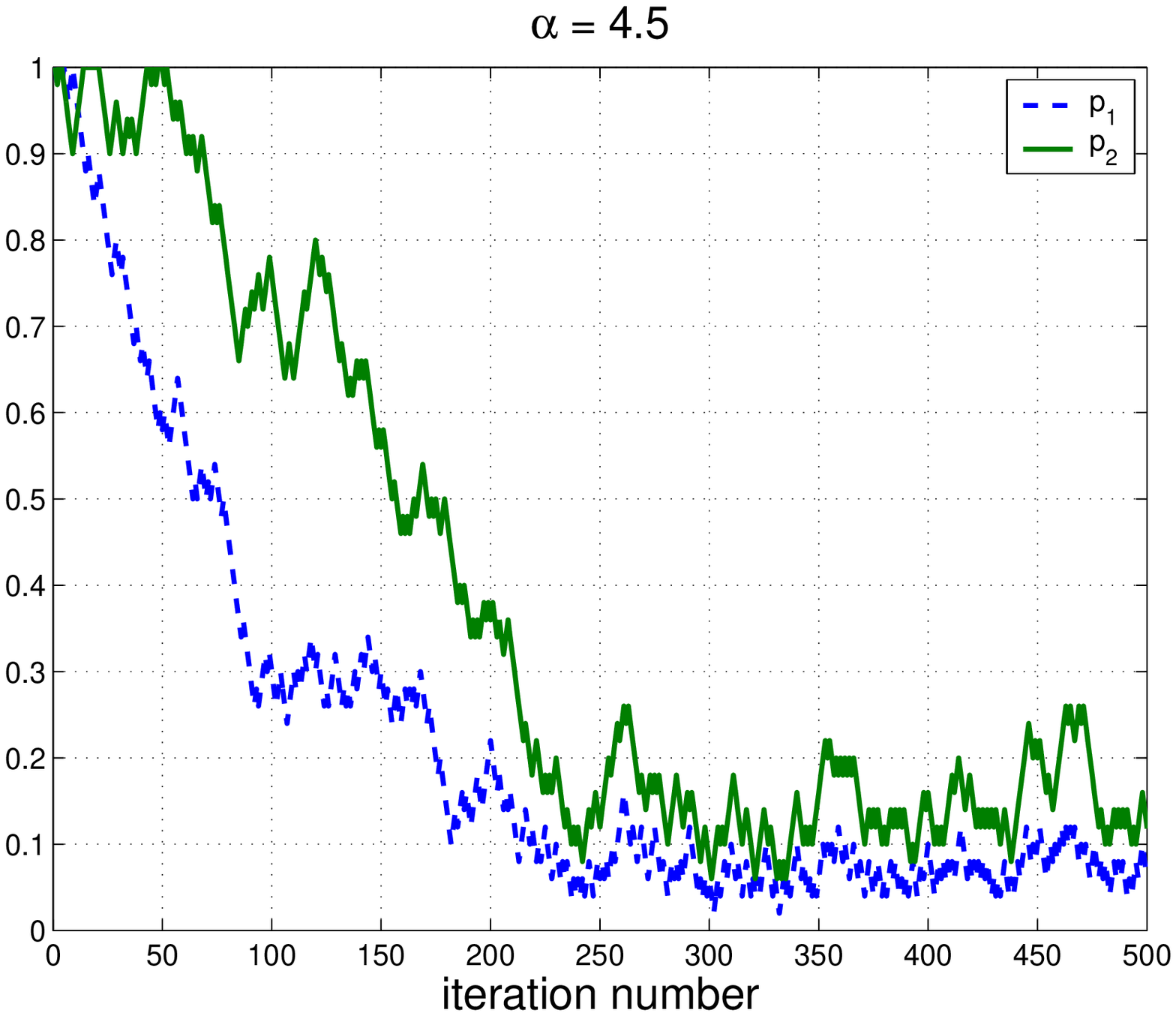}
\caption{Simulations of greedy algorithm under Random Access protocol.}\label{fig:RA_sim}
\end{figure}

The topology used in the simulations consisted of 400 tx nodes from network 1 and 200 tx nodes from network 2, all
i.i.d. uniformly distributed in a square of unit area. For each tx node, its corresponding rx node was located at a
point randomly chosen at uniform from a disc of radius 0.15. This corresponds to $N_1=400/\sqrt(800)\approx 14.14$ and
$N_2=200/\sqrt(800) \approx 7.28$. A step-size of $\Delta = 0.02$ was used. When computing the throughputs, in order to
avoid boundary effects, only transmissions emanating from those tx nodes in the interior of the network were counted.
The results for $\alpha = 2.5, 3.5$ and $4.5$ are displayed in figure \ref{fig:RA_sim}. The observed behavior
corresponds to the analytical results. For the values of $N_1$ and $N_2$ used, the N.E. lies in the Full/Full regime
for $\alpha=2.5$, the Full/Partial regime for $\alpha=3.5$, and the Partial/Partial regime for $\alpha=4.5$, as can be
seen from figure \ref{fig:regimes}.

\subsection{Carrier Sensing Multiple Access based protocol}

The high level conclusion from our analysis of the Random Access protocol, is that the Nash Equilibrium is cooperative
in nature for a sufficiently high pathloss exponent. Ideally we would like to be able to draw this conclusion for a
more sophisticated class of scheduling protocols employing carrier sensing. Due to the analytical intractability of the
problem, we present simulation results to illustrate this effect. We assume both networks operate under the following
protocol. We present a centralized version of it due to space constraints, but claim there exists a distributed version
that performs identically in most cases. During the scheduling phase, each tx-rx pair is assigned a unique token at
random from $\{1,\dots,n\}$. Tx nodes proceed with their transmission so long as they will not cause excessive
interference to any rx node with a higher priority token. More precisely, a transmission is scheduled so long as for
each rx node with higher priority, the difference between its received signal power in dB and the interference power
from the lower priority tx node in dB, exceeds a silencing threshold $\gamma_i$ ($i=1$ for network 1 and $i=2$ for
network 2). Thus a game between the two networks can be defined where the strategies are the choices of silencing
thresholds $\gamma_1$ and $\gamma_2$. We refer to this as the {\it CSMA game}. The silencing threshold for the CSMA
game essentially plays the same role as the access probability in the Random Access game -it determines the degree of
spatial reuse. A high value of $\gamma$ leads to a low density of transmissions, a low value of $\gamma$ leads to a
high density.

We simulate the behavior that arises when both networks optimize their silencing thresholds in a greedy manner.
Analogously to before, we have the following algorithm. \\

\noindent {\bf Inputs:} $p_1(0),p_2(0),\Delta$ \\
{\bf Ouputs:} ${\bf p}_i = [p_i(1),\dots,p_i(500)]$, for $i=1,2$. \\

\noindent {\bf For} $t = 1$ to $500$ \\

Form estimate $\overline{R}_1(\gamma_1(t-1) + \Delta,\gamma_2(t-1))$

Form estimate $\overline{R}_1(\gamma_1(t-1) - \Delta,\gamma_2(t-1))$ \\

{\bf If} $\overline{R}_1(\gamma_1(t-1) + \Delta,\gamma_2(t-1)) > \overline{R}_1(\gamma_1(t-1) - \Delta,\gamma_2(t-1))$

\hspace{3 mm} $\gamma_1(t) =  \min (\gamma_1(t-1) + \Delta,30{\sf dB})$

{\bf Else}

\hspace{3 mm} $\gamma_1(t) = \max (\gamma_1(t-1) - \Delta, -30{\sf dB})$

{\bf End} \\

Form estimate $\overline{R}_2(\gamma_1(t), \gamma_2(t-1) + \Delta)$

Form estimate $\overline{R}_2(\gamma_1(t), \gamma_2(t-1) - \Delta)$ \\

{\bf If} $\overline{R}_2(\gamma_1(t), \gamma_2(t-1) + \Delta) > \overline{R}_2(\gamma_1(t),\gamma_2(t-1) - \Delta)$

\hspace{3 mm} $\gamma_2(t) =  \min( \gamma_2(t-1) + \Delta, 30{\sf dB})$

{\bf Else}

\hspace{3 mm} $\gamma_2(t) = \max(\gamma_2(t-1) - \Delta,-30{\sf dB})$

{\bf End} \\

\noindent {\bf End} \\

The topology used in the setup is identical to before, the only exception being that at each iteration of the
algorithm, 10 old tx-rx pairs leave each network, and 10 new pairs join in i.i.d. locations drawn uniformly at random.
This is to ensure sufficient averaging.

\begin{figure}
\centering
\includegraphics[width=210pt]{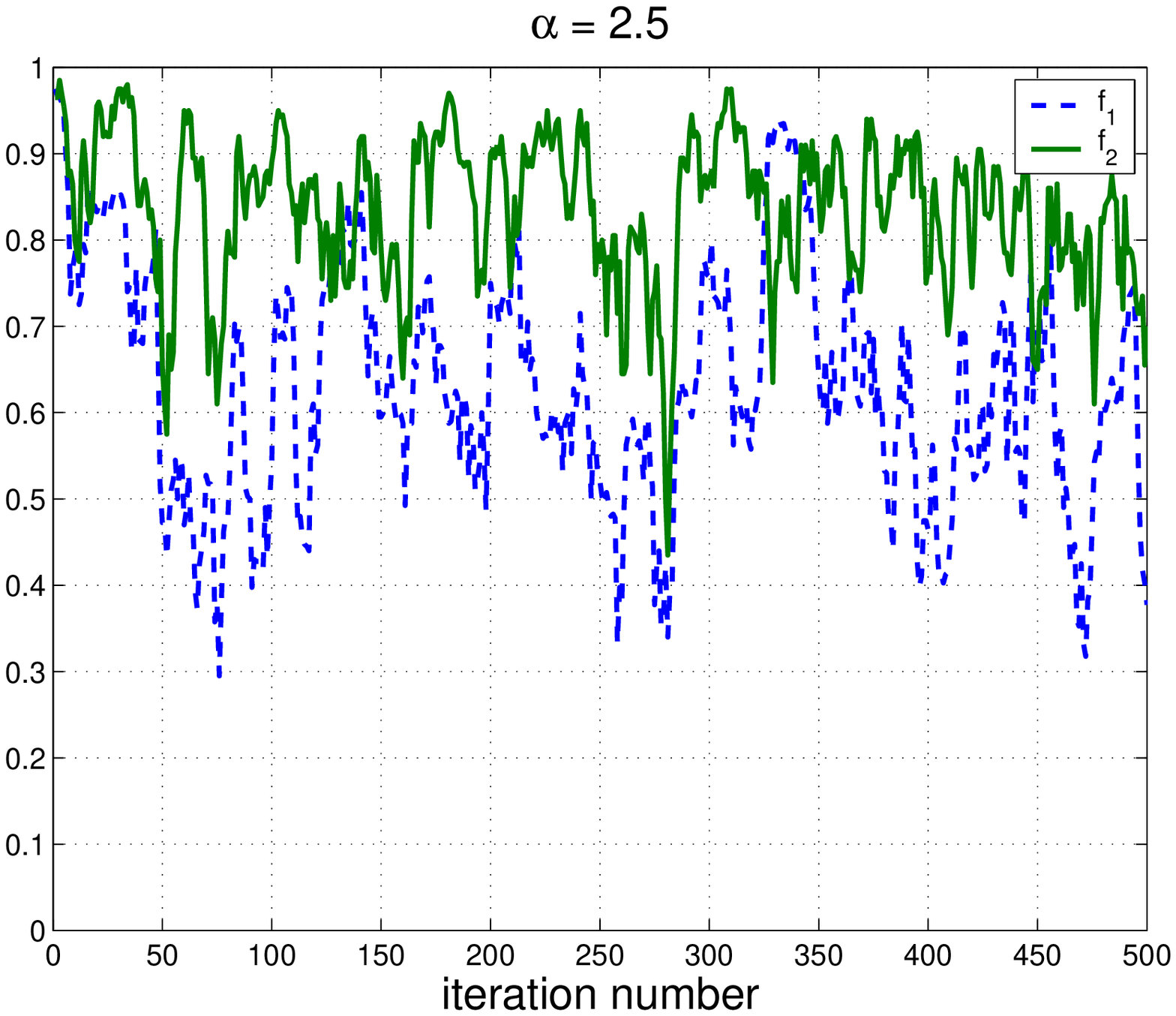}\\
\includegraphics[width=210pt]{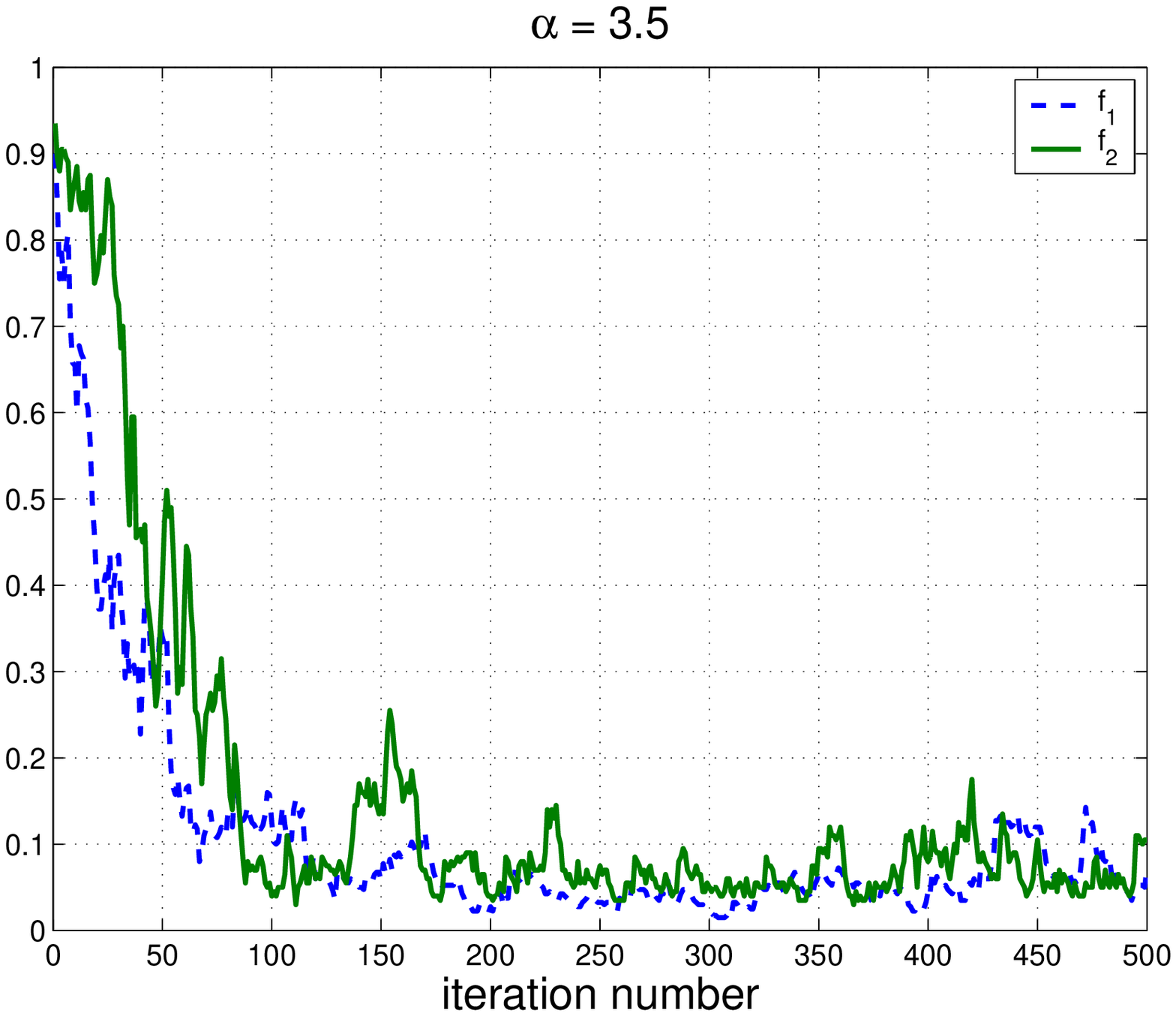}\\
\includegraphics[width=210pt]{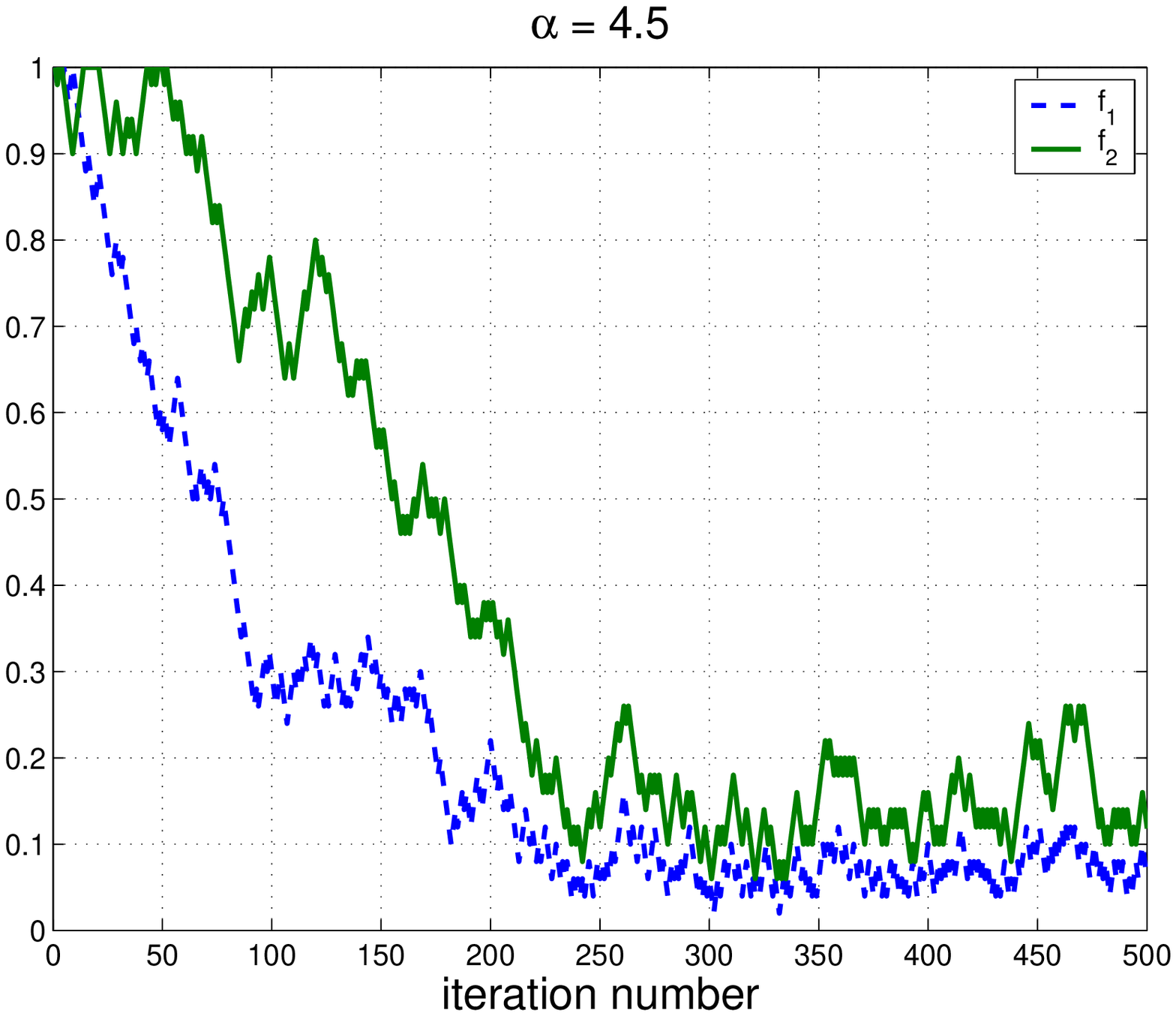}
\caption{Simulations of greedy algorithm under Carrier Sensing Multiple Access protocol}\label{fig:CSMA_sim}
\end{figure}

In a similar fashion to before, each network estimates the effect of either increasing or decreasing the silencing
threshold and then makes a permanent choice. For the same parameter values, the results of the simulation are displayed
in figure \ref{fig:CSMA_sim}. On the y-axes of these plots we have drawn the fraction of nodes simultaneously scheduled
at each iteration, which we denote $f_1$ and $f_2$, rather than the silencing thresholds $\gamma_i$, in order to draw a
simple visual comparison with figure \ref{fig:RA_sim}. For this reason there is more fluctuation in the results, as the
fraction of simultaneously scheduled transmissions varies not only due to the up/down movements of the silencing
thresholds, but also due to the changing topology.

We conclude from these plots that for small values of $\alpha$ (namely $\alpha=2.5$) the system converges to a
competitive equilibrium, where both networks simultaneously schedule a large fraction of their transmissions, and for
large values of $\alpha$ (namely $\alpha=3.5$ and $4.5$) the system converges to a near cooperative equilibrium, where
both networks schedule a small fraction of their transmissions.

\section{Conclusion}
This work studied spectrum sharing between wireless devices operating under a random access protocol. The crucial
assumption made was that nodes belonging to the same network or coalition cooperate with one another. Competition only
exists between nodes belonging to rival networks. It was found that cooperation between devices within each network
created the necessary incentive to prevent total anarchy. For pathloss exponents greater than four, we showed that
contrary to ones intuition, there can be a natural incentive for devices to cooperate to the extent that each occupies
only a fraction of the available bandwidth. Such results are optimistic and encouraging. We demonstrated via
simulations that it may be possible to extend them to more complex operating protocols such as those that employ
carrier-sensing to determine when the medium is free. More generally one wonders whether a multi-stage game capturing
the system dynamics under such a protocol can be formulated, and whether the desirable properties of the single-stage
game continue to hold. It would also be worthwhile investigating the incentives wireless links have to form coalitions,
as in this work it was in essence assumed that coalitions had been pre-determined.

\section{Proofs}
\subsection{Theorem \ref{thm:op_access_prob}}
The limiting expression for the average throughput is
\begin{equation*}
{\mathbb E}{\overline R}(p_i,\beta_i) \rightarrow p_i \log(1+\beta_i) e^{-N_ip_i\beta_i^{2/\alpha}}.
\end{equation*}

\noindent Given a $\beta_i$ there is a single maximum over $p_i$. By differentiation we have
\begin{equation*}
\frac{\partial {\mathbb E}{\overline R}}{\partial{p_i}} = \left(1-p_iN_i\beta_i^{2/\alpha}\right)
\log(1+\beta_i)e^{-N_ip_i\beta_i^{2/\alpha}},
\end{equation*}

\noindent so $p_i^* = \min(1,1/N_i\beta_i^{2/\alpha})$. Thus
\begin{equation*}
{\mathbb E}{\overline R}(p_i^*,\beta_i) = \left\{
                             \begin{array}{ll}
                               \log(1+\beta_i)e^{-N_i\beta_i^{2/\alpha}}, &\hbox{$\beta_i\le N_i^{-2/\alpha}$}, \\
                               \dfrac{\log(1+\beta_i)e^{-1}}{N_i\beta_i^{2/\alpha}}, &\hbox{$\beta_i > N_i^{-2/\alpha}$}.
                             \end{array}
                           \right.
\end{equation*}

\noindent Both of these functions have one maximum, but the maximum of the second function is always greater than the
maximum of the first as it represents the solution to the unconstrained problem
\begin{equation*}
\max_{p_i>0,\beta_i>0}{\mathbb E}{\overline R}(p_i,\beta_i),
\end{equation*}

\noindent whereas the maximum of the first represents the solution to
\begin{equation*}
\max_{p_i=1,\beta_i>0}{\mathbb E}{\overline R}(p_i,\beta_i).
\end{equation*}

\noindent Thus if the maximum of the second function occurs for $\beta_i > N_i^{-2/\alpha}$, it is the maximum of the
entire function, but if it occurs for $\beta_i \le N^{-2/\alpha}$, the maximum of the entire function is the maximum of
the first function over the domain $\beta_i \le N^{-2/\alpha}$. By differentiation we find the maximum of the second
function occurs at the unique solution of
\begin{equation*}
\frac{\alpha}{2} = (1+1/\beta_i)\log(1+\beta_i)
\end{equation*}

\noindent which is $\beta_i={\Lambda^*}^{-\alpha/2}$. Thus if $N_i>\Lambda^*$, the solution is $p^* = \Lambda^*/N_i$
and $\beta_i^* = {\Lambda^*}^{-\alpha/2}$. If $N_i\le \Lambda^*$ the solution is $p^* = 1$ and $\beta_i^*$ equal to the
unique solution of
\begin{equation}\label{eqn:2ndfn_max}
\frac{\alpha}{2N_i\beta_i^{2/\alpha}} = (1+1/\beta_i)\log(1+\beta_i)
\end{equation}

\noindent or $N_i^{-2/\alpha}$, whichever is smaller. But as $(1+1/x)\log(1+x)$ is a monotonically increasing function,
from the definition of $\Lambda^*$ we have
\begin{equation*}
\frac{\alpha}{2} \le \left(1+N_i^{\alpha/2}\right) \log \left( 1 + N_i^{-\alpha/2} \right)
\end{equation*}

\noindent whenever $N_i\le \Lambda^*$. Combining this with equation (\ref{eqn:2ndfn_max}) and using the monotonicity of
$(1+1/x)\log(1+x)$ we see that the unique solution to equation (\ref{eqn:2ndfn_max}) is always smaller than
$N_i^{-2/\alpha}$. This establishes the desired result.

\subsection{Theorem \ref{thm:opt_p2}}
First we show that any N.E. must lie on the boundary of the strategy space, i.e. $\Lambda_i=1$ for some $i$. The
utility functions are smooth and continuous. Differentiating $U_1$ with respect to $\Lambda_1$ yields
\begin{equation*}
\frac{\partial U_1}{\partial \Lambda_1} = e^{-\Lambda_2/\Lambda_1}\left(
\frac{\Lambda_2}{\Lambda_1}-\frac{\alpha}{2\left( 1 + \Lambda_1^{\alpha/2}\right)
\log\left(1+\Lambda_1^{-\alpha/2}\right)} + 1\right)
\end{equation*}

\noindent Consider the function $f(x)=(1+x)\log(1+1/x)$. As $f(x)$ is monotonically decreasing for $x\ge 0$ and
$\lim_{x\rightarrow \infty} f(x) = 1$ we have $f(x)> 1$ for all $x\ge 0$. Thus for $\alpha<4$, ${\partial
U_1}/{\partial \Lambda_1}
> 0$ whenever $\Lambda_1<\Lambda_2$ and similarly ${\partial U_2}/{\partial \Lambda_2} > 0$ whenever $\Lambda_2<\Lambda_1$. Thus for a N.E. to occur
in the interior of the strategy space we must have both $\Lambda_1>\Lambda_2$ and $\Lambda_2>\Lambda_1$. As these
conditions are mutually exclusive at least one of the constraints of the strategy space must be active at the N.E.. In
essence each network is trying to set it's active density higher than the other's. Eventually at least one network maxs
out.

First consider the case where the solution to
\begin{equation}\label{eqn:opt_p2_again}
N_1 = x\left(\frac{\alpha}{2\left(1+{x}^{\alpha/2}\right)\log \left(1+{x}^{-\alpha/2}\right)} - 1\right).
\end{equation}

\noindent occurs for $x < N_2$. Suppose $\Lambda_2^* = N_2$. Then as $N_1 \le N_2$ we have $\Lambda_1\le \Lambda_2^*$,
hence $\partial U_1/\partial \Lambda_1 > 0$ for all $\Lambda_1$ on the interior and hence $\Lambda_1^* = N_1$. Now the
function $U_2(\Lambda_1^*,\Lambda_2)$ has a unique maximum for $\Lambda_2$. This maximum satisfies $\partial
U_2(\Lambda_1^*,\Lambda_2)\partial \Lambda_2=0$, which is equation (\ref{eqn:opt_p2_again}) with $\Lambda_2^*$
substituted for $x$. But the solution equation (\ref{eqn:opt_p2_again}) satisfies $x < N_2$, so $\Lambda_2^* < N_2$, a
contradiction. Thus the constraint $\Lambda_2 \le N_2$ must be inactive. Suppose instead that the constraint
$\Lambda_1^* = N_1$ is active. Then by the same arguments the unique $\Lambda_2^*$ satisfies equation
(\ref{eqn:opt_p2_again}). This establishes the solution and it's uniqueness, for the first case.

Second consider the case where the solution to equation (\ref{eqn:opt_p2_again}) occurs for $x\ge N_2$. Then it is
straightforward to check using similar arguments above, that the unique solution satisfies
$(\Lambda_1^*,\Lambda_2^*)=(N_1,N_2)$. This establishes the result.

\subsection{Theorem \ref{thm:asymp_soln}}
For $\alpha>2$ the solution to equation (\ref{eqn:opt_p2}) only goes to infinity for $N_1\rightarrow \infty$. In this
limit $(1+{\Lambda_2^*}^{\alpha/2})\log(1+{\Lambda_2^*}^{-\alpha/2})\rightarrow 1$ and
\begin{equation*}
\Lambda_2^* \rightarrow \frac{2}{\alpha-2}\Lambda_1^*
\end{equation*}

\noindent if $N_1\le (\alpha/2-1)N_2$. Otherwise, $\Lambda_2^*=N_2$. Computing $p_i^*=\Lambda_i^*/N_i$ produces the
stated result.

\subsection{Theorem \ref{thm:asymp_util}}
This follows by direct substitution.

\subsection{Theorem \ref{thm:a_ge_4}}
The proof of this result is more involved than the proof of theorem 2.3. There are three regimes.

First consider the joint spread regime where $N_1 > \sqrt{\Lambda^*(\alpha/2)}$. We show that a N.E. cannot occur on
the boundary of the strategy space. Suppose $\Lambda_1^*=N_1$. Then $\Lambda_1^*>\sqrt{\Lambda^*(\alpha/2)}$. As
$\sqrt{\Lambda^*(\alpha/2)}$ is the solution to the equation
\begin{equation*}
\frac{\alpha}{2\left( 1 + {\sqrt{\Lambda^*(\alpha/2)}}^{\alpha/2}\right) \log \left( 1 +
{\sqrt{\Lambda^*(\alpha/2)}}^{-\alpha/2} \right)}- 1 = 1.
\end{equation*}

\noindent and $(1+x)\log(1+1/x)$ is a monotonically decreasing function for $x>0$, we have
\begin{equation*}
\frac{\alpha}{2\left( 1 + {\Lambda_1^*}^{\alpha/2}\right) \log \left( 1 + {\Lambda_1^*}^{-\alpha/2} \right)} - 1 > 1.
\end{equation*}

\noindent If $\Lambda_1^*$ lies on the boundary of the strategy space then $\partial U_1(\Lambda_1^*,\Lambda_2^*)
/\partial \Lambda_1 > 0$ which implies
\begin{align*}
\Lambda_2^* &> \Lambda_1^*\left( \frac{\alpha}{2\left( 1 + {\Lambda_1^*}^{\alpha/2}\right) \log \left( 1 +
{\Lambda_1^*}^{-\alpha/2} \right)} -
1\right) \\
&> \Lambda_1^* \\
&> \sqrt{\Lambda^*(\alpha/2)}.
\end{align*}

\noindent This in turn implies
\begin{equation*}
\frac{\alpha}{2\left( 1 + {\Lambda_2^*}^{\alpha/2}\right) \log \left( 1 + {\Lambda_2^*}^{-\alpha/2} \right)} - 1 > 1.
\end{equation*}

\noindent At equilibrium $\partial U_2(\Lambda_1^*,\Lambda_2) / \partial \Lambda_2 \ge 0$ so
\begin{align*}
\Lambda_1^* &\ge \Lambda_2^*\left( \frac{\alpha}{2\left( 1 + {\Lambda_2^*}^{\alpha/2}\right) \log \left( 1 +
{\Lambda_2^*}^{-\alpha/2} \right)} -
1\right) \\
&> \Lambda_2^*,
\end{align*}

\noindent a contradiction. Thus $\Lambda_1 < N_1$. Now assume $\Lambda_2 = N_2$. By assumption $N_2 \ge N_1$ so
$\Lambda_2^*
> \sqrt{\Lambda^*(\alpha/2)}$. By repeating the same arguments we can generate the same style of contradiction and thus $\Lambda_2
< N_2$. This establishes that a N.E. cannot occur on the boundary of the strategy space. In essence each network is
trying to undercut the active density of the other. This drags the equilibrium away from the boundary.

Now we establish any N.E. must be symmetric, i.e. $\Lambda_1^*=\Lambda_2^*$. Suppose a N.E. $(\Lambda_1^*,\Lambda_2^*)$
with $\Lambda_1^* \neq \Lambda_2^*$ exists. Then as it must lie on the interior of the strategy space and as the
utility functions are symmetric, $(\Lambda_2^*,\Lambda_1^*)$ must also be a N.E.. On the interior of the strategy space
the N.E. criterion is $\partial U_1(\Lambda_1^*,\Lambda_2^*) /\partial \Lambda_1 = 0$ and so the function
$\Lambda_1^*(\Lambda_2)$ is monotonically increasing in $\Lambda_2$. But this implies we cannot have N.E. at both
$(\Lambda_1^*,\Lambda_2^*)$ and $(\Lambda_2^*,\Lambda_1^*)$, a contradiction. Thus $\Lambda_1^*=\Lambda_2^*$.

By differentiating the utility functions this implies that at any N.E. $\Lambda_1^*$ satisfies
\begin{equation*}
\frac{\alpha}{2\left( 1 + {\Lambda_1^*}^{\alpha/2}\right) \log \left( 1 + {\Lambda_1^*}^{-\alpha/2} \right)} - 1 = 1
\end{equation*}

\noindent with $\Lambda_2^* = \Lambda_1^*$. But this is equivalent to $\Lambda_1^* = \sqrt{\Lambda^*(\alpha/2)}$. Thus
the N.E. is unique and occurs at $(\Lambda_1^*,\Lambda_2^*) = (\sqrt{\Lambda^*(\alpha/2)},\sqrt{\Lambda^*(\alpha/2)})$.

Next consider the partial spread and full spread regimes where $N_1 \le \sqrt{\Lambda^*(\alpha/2)}$. We first show that
$\Lambda_1^* = N_1$. Suppose $\Lambda_1^* < N_1$. Then $\Lambda_1^* < \sqrt{\Lambda^*(\alpha/2)}$ which implies
\begin{equation*}
\frac{\alpha}{2\left( 1 + {\Lambda_1^*}^{\alpha/2}\right) \log \left( 1 + {\Lambda_1^*}^{-\alpha/2} \right)} - 1 < 1.
\end{equation*}

\noindent At equilibrium $\partial U_1/\partial \Lambda_1 = 0$ so
\begin{align*}
\Lambda_2^* &= \Lambda_1^*\left( \frac{\alpha}{2\left( 1 + {\Lambda_1^*}^{\alpha/2}\right) \log \left( 1 +
{\Lambda_1^*}^{-\alpha/2} \right)} -
1\right) \\
&< \Lambda_1^* \\
&< \sqrt{\Lambda^*(\alpha/2)}.
\end{align*}

\noindent But this in turn implies
\begin{equation*}
\frac{\alpha}{2\left( 1 + {\Lambda_2^*}^{\alpha/2}\right) \log \left( 1 + {\Lambda_2^*}^{-\alpha/2} \right)} - 1 < 1
\end{equation*}

\noindent which in conjunction with the equilibrium condition $\partial U_2/\partial \Lambda_2 = 0$ implies
$\Lambda_1^*<\Lambda_2^*$, a contradiction. Thus $\Lambda_1^* = N_1$. Now we can solve for $\Lambda_2^*$ to conclude
that $\Lambda_2^*$ is the unique solution to equation (\ref{eqn:opt_p2}) or $N_2$, whichever is smaller. This concludes
the proof.

\subsection{Theorem \ref{thm:vr_soln}}

We first tackle the full spread and partial spread regimes. It is shown in lemma \ref{lem:4} in the appendix that
$\Lambda''(\alpha)$ is undefined for $\alpha\le 4$ and recall $\Lambda''(\alpha)$ is defined to be positive infinity
for $\alpha > 4$. Consider the case where $N_1\le \Lambda''(\alpha)$. We show that $\Lambda_1^*=N_1$. Suppose the
contrary, that $\Lambda_1^*<N_1$. Then $\Lambda_1^* < \Lambda''(\alpha)$. Define the function
\begin{equation*}
f(s_1,s_2) \triangleq \dfrac{\int_0^{\infty} e^{-(s_1+s_2)x^{2/\alpha}} \dfrac{dx}{1+x}}{s_1\int_0^{\infty}
x^{2/\alpha}e^{-(s_1+s_2)x^{2/\alpha}} \dfrac{dx}{1+x} },
\end{equation*}

\noindent for $s_1$ and $s_2$ positive. In Lemma \ref{lem:1} in the appendix it is shown that $f(s,s)$ is a
monotonically decreasing function in $s$. By rearranging equation (\ref{eqn:Lambda_doubleprime}) one can check that
$f(\Lambda''(\alpha),\Lambda''(\alpha))=1$. Thus $f(\Lambda_1^*,\Lambda_1^*)>1$. For a given $\Lambda_2$ it is shown in
Lemma \ref{lem:2} in the appendix that the utility function $U_1(\Lambda_1,\Lambda_2)$ is a smooth continuous function
of $\Lambda_1$ with a unique maximum (and likewise for $U_2$ given $\Lambda_1$). As $\Lambda_1^*<N_1$ the maximum with
respect to $\Lambda_1$ occurs at $\partial U_1(\Lambda_1^*,\Lambda_2)/\partial \Lambda_1 = 0$. By rearranging equation
(\ref{eqn:vr_thm}) one can check that this condition is equivalent to $f(\Lambda_1^*,\Lambda_2)=1$. This means
$f(\Lambda_1^*,\Lambda_2^*) < f(\Lambda_1^*,\Lambda_1^*)$. In lemma \ref{lem:2} we show that $f(s_1,s_2)$ is a
monotonically increasing function in $s_2$ given $s_1$. Thus we have $\Lambda_2^*<\Lambda_1^*$ and so also
$\Lambda_2^*<\Lambda''(\alpha)$ and $\Lambda_2^*<N_1$. Thus $f(\Lambda_2^*,\Lambda_2^*)>1$. As by assumption $N_1\le
N_2$, we have $\Lambda_2^*<N_2$ and so the maximum of $U_2$ occurs at $\partial U_2(\Lambda_1,\Lambda_2^*)/\partial
\Lambda_2 = 0$. This means $f(\Lambda_2^*,\Lambda_1^*) = 1 < f(\Lambda_2^*,\Lambda_2^*)$ which implies
$\Lambda_1^*<\Lambda_2^*$. This is a contradiction. Thus we must have $\Lambda_1^* = N_1$ at a N.E.. By maximizing over
$\Lambda_2$ via differentiation of $U_2$, we see that $\Lambda_2^*$ equals the solution of (\ref{eqn:vr_thm}) or $N_2$
whichever is smaller. Lemma \ref{lem:2} establishes the solution of (\ref{eqn:vr_thm}) always exists and is unique.

Now consider the case where $N_1 > \Lambda''(\alpha)$. We first show that a N.E. cannot occur on the boundary of the
strategy space. Suppose $\Lambda_1^*=N_1$. Then $\Lambda_1 > \Lambda''(\alpha)$. This implies
$f(\Lambda_1^*,\Lambda_1^*) <1$. As $\Lambda_1^*=N_1$ the maximum of $U_1$ occurs at $\partial
U_1(\Lambda_1^*,\Lambda_2)/\partial \Lambda_1 \ge 0$ for a given $\Lambda_2$. This means $f(\Lambda_1^*,\Lambda_2^*)
\ge 1 > f(\Lambda_1^*,\Lambda_1^*)$, thus $\Lambda_2^* > \Lambda_1^*$. We also then have $\Lambda_2^*>
\Lambda''(\alpha)$. From the optimality condition for network 2 we then have $\partial
U_2(\Lambda_1^*,\Lambda_2)/\partial \Lambda_2\ge 0$. This means $f(\Lambda_2^*,\Lambda_1^*) \ge 1 >
f(\Lambda_2^*,\Lambda_2^*)$ which implies $\Lambda_1^*
> \Lambda_2^*$. This is a contradiction. Thus we must have $\Lambda_1^* < N_1$. As $N_2\ge N_1>\Lambda''(\alpha)$ we can repeat the
argument for $\Lambda_2^*$ to conclude that we must also have $\Lambda_2^*< N_2$. This proves a N.E. can only occur on
the interior of the strategy space.

Now we establish any N.E. must be symmetric, i.e. $\Lambda_1^*=\Lambda_2^*$. Suppose a N.E. $(\Lambda_1^*,\Lambda_2^*)$
with $\Lambda_1^* \neq \Lambda_2^*$ exists. Then as it must lie on the interior of the strategy space and as the
utility functions are symmetric, $(\Lambda_2^*,\Lambda_1^*)$ must also be a N.E.. On the interior of the strategy space
the N.E. criterion is $\partial U_1(\Lambda_1^*,\Lambda_2^*) /\partial \Lambda_1 = 0$ and so the function
$\Lambda_1^*(\Lambda_2)$ is monotonically increasing in $\Lambda_2$ by lemma \ref{lem:3}. But this implies we cannot
have N.E. at both $(\Lambda_1^*,\Lambda_2^*)$ and $(\Lambda_2^*,\Lambda_1^*)$, a contradiction. Thus
$\Lambda_1^*=\Lambda_2^*$.

Finally one can verify that $\Lambda_1^*=\Lambda_2^*=\Lambda''(\alpha)$ is a N.E. by differentiating the utility
functions. Thus the unique N.E. is $(\Lambda_1^*,\Lambda_2^*) = (\Lambda''(\alpha),\Lambda''(\alpha))$.

In essence what is going on here is that in the absence of strategy space constraints, when
$\Lambda_i<\Lambda''(\alpha)$, network $i$ wants to set $\Lambda_i>\Lambda_j$ and when $\Lambda_i>\Lambda''(\alpha)$
network $i$ wants to set $\Lambda_i<\Lambda_j$. Thus the natural equilibrium is at $(\Lambda_1^*,\Lambda_2^*) =
(\Lambda''(\alpha),\Lambda''(\alpha))$. The problem for $\alpha<4$ is $\Lambda''(\alpha)$ is infinite and the sparser
network winds up maxing out at $\Lambda_1^*=N_1$. When $\alpha>4$ the function $\Lambda''(\alpha)$ is finite and it is
possible to have $N_1>\Lambda''(\alpha)$, i.e. both networks have a sufficiently high density of nodes so as not to be
constrained by the strategy space. In this case they get to set their access probabilities so as to achieve the natural
equilibrium.

\section{Appendix}
\begin{lemma}\label{lem:1}
The function $f(s,s)$ is monotonically decreasing in $s$.
\end{lemma}
\begin{proof}
By changing variables we can rewrite $f(s,s)$ as
\begin{equation*}
f(s,s) \triangleq \dfrac{\int_0^{\infty}g_s(x)dx}{\int_0^{\infty}xg_s(x)dx}
\end{equation*}

\noindent where
\begin{equation*}
g_s(x) = \dfrac{x^{\alpha/2-1}e^{-2x}}{s^{\alpha/2}+x^{\alpha/2}}.
\end{equation*}

\noindent Choose a pair of values for $s_1$ and $s_2$ satisfying $0\le s_1\le s_2$ and then observe that for any
$x_1\le x_2$ the following inequality holds
\begin{equation*}
\dfrac{g_{s_2}(x_2)}{g_{s_1}(x_2)} \ge \dfrac{g_{s_2}(x_1)}{g_{s_1}(x_1)}.
\end{equation*}

\noindent Thus
\begin{align*}
\int_0^{\infty}\int_0^{x_2} (x_2-x_1)g_{s_1}(x_1)g_{s_2}(x_2)dx_1dx_2 &= \int_0^{\infty}\int_{x_1}^{\infty} (x_2-x_1)g_{s_1}(x_1)g_{s_2}(x_2)dx_2dx_1 \\
&= \int_0^{\infty}\int_{x_2}^{\infty} (x_1-x_2)g_{s_1}(x_2)g_{s_2}(x_1)dx_1dx_2 \\
&\ge \int_0^{\infty}\int_{x_2}^{\infty} (x_1-x_2)g_{s_1}(x_1)g_{s_2}(x_2)dx_1dx_2.
\end{align*}

\noindent Then
\begin{equation*}
\int_0^{\infty}\int_0^{\infty} (x_2-x_1)g_{s_1}(x_1)g_{s_2}(x_2)dx_1dx_2 \ge 0
\end{equation*}

\noindent and so
\begin{equation*}
\int_0^{\infty}\int_0^{\infty} x_2g_{s_1}(x_1)g_{s_2}(x_2)dx_1dx_2 \ge \int_0^{\infty}\int_0^{\infty}
x_1g_{s_1}(x_1)g_{s_2}(x_2)dx_1dx_2
\end{equation*}

\noindent which implies
\begin{equation*}
\int_0^{\infty} g_{s_1}(x)dx \int_0^{\infty} xg_{s_2}(x)dx \ge \int_0^{\infty} xg_{s_1}(x)dx \int_0^{\infty}
g_{s_2}(x)dx
\end{equation*}

\noindent and thus $f(s_1,s_1)\ge f(s_2,s_2)$.
\end{proof}

\begin{lemma}\label{lem:2}
The function
\begin{equation*}
U_i(\Lambda_1,\Lambda_2) = \Lambda_i\int_0^{\infty} e^{-(\Lambda_1+\Lambda_2)x^{2/\alpha}}\dfrac{dx}{1+x}
\end{equation*}

\noindent is smooth and continuous in $\Lambda_i$ with a unique maximum $\Lambda_i^*$.
\end{lemma}

\begin{proof}
As the integral is well-defined for all positive $\Lambda_1$ and $\Lambda_2$, the function is smooth and continuous by
inspection. To see that a unique maximum exists set the derivative to zero to obtain $f(\Lambda_i,\Lambda_j) = 1$. For
fixed $\Lambda_j$ it is straightforward to show $f(\Lambda_i,\Lambda_j)$ is monotonically decreasing in $\Lambda_i$
using arguments similar to those in lemma \ref{lem:1}. For $\Lambda_i\rightarrow 0$ we find
$f(\Lambda_i,\Lambda_j)\rightarrow\infty$ and for $\Lambda_i \rightarrow \infty$ we find $f(\Lambda_i,\Lambda_j)
\rightarrow 2/\alpha$ which is always less than 1 for $\alpha>2$. Thus there always exists a single $\Lambda_i^*$
satisfying $f(\Lambda_i^*,\Lambda_j)=1$ and hence a unique maximum always exists.
\end{proof}

\begin{lemma}\label{lem:3}
The function $f(s_1,s_2)$ is monotonically increasing in $s_2$ given $s_1$.
\end{lemma}

\begin{proof}
The proof mirrors that of lemma \ref{lem:1}, the only difference being that now
\begin{equation*}
\dfrac{g_{s_2}(x_2)}{g_{s_1}(x_2)} \le \dfrac{g_{s_2}(x_1)}{g_{s_1}(x_1)},
\end{equation*}

\noindent i.e. the inequality goes the other way. We omit the details for brevity.
\end{proof}

\begin{lemma}\label{lem:4}
The function $\Lambda''(\alpha)$ is undefined for $\alpha \le 4$ and uniquely defined for $\alpha > 4$.
\end{lemma}

\begin{proof}
$\Lambda''(\alpha)$ is the solution to $f(\Lambda''(\alpha),\Lambda''(\alpha))=1$. The function $f(s,s)$ is a
monotonically decreasing in $s$ by lemma \ref{lem:1}. By taking $s\rightarrow 0$ we find $f(s,s)\rightarrow \infty$ and
by taking $s\rightarrow \infty$ we find $f(s,s)\rightarrow 4/\alpha$. Thus when $\alpha\le 4$ there is no $s$ for which
$f(s,s)=1$ and hence $\Lambda''(\alpha)$ is undefined. When $\alpha > 4$ there is a single $s$ at which $f(s,s)$
crosses the value 1 and hence $\Lambda''(\alpha)$ is uniquely defined.
\end{proof}

\end{document}